\documentclass[12pt]{article}
\usepackage[a4paper,textwidth=15.5cm,textheight=22cm,footnotesep=\baselineskip]{geometry}
\usepackage{bm}      
\usepackage{amsmath} 
\usepackage{cite}    
\usepackage{slashed}
\usepackage{hyperref}
\hypersetup{final}
\usepackage{bm}
\usepackage[utf8]{inputenc}
\usepackage{mathrsfs} 

\usepackage[all]{nowidow}

\usepackage{graphicx}

\usepackage[all]{hypcap}

\def\lsim{\mbox{~{\raisebox{0.4ex}{$<$}}\hspace{-1.1em}
        {\raisebox{-0.6ex}{$\sim$}}~}}

\renewcommand{\vec}[1]{{\bf #1}}
\newcommand*\GammaLL  \Gamma 
\newcommand*\gammaLL { \gamma } 
\newcommand*\LeR { L _ {  e  { \rm R }}} 
\newcommand*\laplace{\mathop{}\!\mathbin\bigtriangleup}

\begin{document}
\setlength{\baselineskip}{0.6cm} 

\makeatletter \@addtoreset{equation}{section} \makeatother
\renewcommand{\theequation}{\arabic{section}.\arabic{equation}}

\newcommand*\xbar[1]{%
  \hbox{%
    \vbox{%
      \hrule height 0.5pt 
      \kern0.5ex
      \hbox{%
        \kern-0.1em
        \ensuremath{#1}%
        \kern-0.1em
      }%
    }%
  }%
}

\begin{centering}

  \textbf{\Large{Equilibration of 
      right-handed electrons }}

\vspace*{.6cm}

Dietrich~B\"odeker
\footnote{bodeker@physik.uni-bielefeld.de%
} and Dennis~Schr\"oder
\footnote{dennis@physik.uni-bielefeld.de%
}

\vspace*{.6cm} 

{\em 
  Fakult\"at f\"ur Physik,
  Universit\"at Bielefeld,
  33501 Bielefeld,
  Germany
}

\vspace{10mm}

\end{centering}

\vspace{5mm}
\noindent

\begin{abstract}
\noindent
  We study the equilibration of right-handed electrons
  in the symmetric phase of the Standard Model. 
  Due to the smallness of the electron Yukawa coupling, it 
  happens relatively late in the history of the Universe. 
  We compute the equilibration rate at leading order 
  in the Standard Model couplings,
  by including gauge interactions, the top Yukawa- and the 
  Higgs self-interaction. The dominant contribution
  is due to $ 2 \to 2 $ particle scattering, even though the 
  rate of (inverse)  Higgs decays is strongly enhanced by
  multiple soft scattering which is included by Landau-Pomeranchuk-Migdal
  (LPM) resummation. Our numerical result is substantially larger than 
  approximations
  presented in previous literature.
\end{abstract}

\section {Introduction}
\label{s:intro}

The electron Yukawa coupling is the smallest coupling constant of the
Standard Model. Therefore thermal  equilibrium between 
right- and left-handed electrons is achieved
rel\-a\-tively late in the evolution of the Universe. Nevertheless, it happened
in the symmetric phase, 
while electroweak sphaleron processes were still rapidly violating baryon plus
lepton number.
Therefore the equilibration of right-handed electrons can play an 
important role in 
the creation of the matter- antimatter-asymmetry
of the Universe.

A matter-antimatter asymmetry created
at some very high temperature like, {\it  e.g.} in GUT baryogenesis,
can be protected from washout if the right-handed
electrons \cite{Campbell:1992jd,Cline:1993bd} are not yet in
equilibrium.  Baryogenesis through neutrino oscillations
\cite{Akhmedov:1998qx,Asaka:2005pn},  for  certain model parameters,
can take place at the same time as the 
the equilibration of the right-handed electrons.
Then the latter is part of the leptogenesis process.
A lepton asymmetry in right-handed electrons 
may also generate hypermagnetic fields \cite{Joyce:1997uy}. 

The importance of electron equilibration was first pointed
out in \cite{Campbell:1992jd}, where it was noted
that the final baryon asymmetry is exponentially sensitive
to the equilibration rate. A computation in \cite{Campbell:1992jd}
included only the inverse Higgs decay. The importance of 
$ 2 \to 2 $ scattering was noted in \cite{Cline:1993bd}. 
The equilibration of heavier 
lepton flavors in thermal leptogenesis was studied in
\cite{Beneke:2010dz,Garbrecht:2013urw}. It was pointed out that
multiple soft $ 1n \leftrightarrow  2 n $
scattering processes also contribute at leading order
\cite{Garbrecht:2013urw}, and the corresponding rate was estimated,
but it has not been computed so far.

In this paper we only consider the equilibration rate
of right-handed  electrons  $ e _ { { \rm R } } $.
We improve on previous calculations 
by correctly treating various thermal effects. 
Furthermore, for the first time, we compute the contributions from multiple
soft gauge interactions in collinear emission processes. 
We consider
temperatures $ T $ well below $ 10 ^{ 14 } $
GeV. Then  weak hypercharge 
interactions are much faster than the Hubble expansion,
and the right-handed electrons are in 
kinetic equilibrium. 
Due to the smallness of the electron Yukawa coupling the lepton
number $ L _ { e { \rm R } } $ carried by $ e _ { \rm R } $ 
takes much longer to equilibrate. For sufficiently small deviations
from equilibrium the time 
evolution of $ L _ { e { \rm R } } $ 
can be described by a linear  
equation. 
Without Hubble expansion it can be written as 
\begin{align}
  \dot L _ { e { \rm R } } 
   = 
  - \gammaLL
  \left ( 
   L _ { e { \rm R } } - L _ { e { \rm R } }  ^{  \rm eq } 
   \right ) 
  + \cdots 
   .
\label{kineqn}
\end{align}
There may by additional contributions on the right-hand side due to
other slowly varying charges. 
Furthermore, the chiral anomaly violates
$ \LeR $-conservation in the Standard Model. Therefore  $ \LeR $ can 
be converted into hypercharge
electromagnetic fields, changing the value of $ \LeR $.
In the absence of long-range gauge fields this is a non-linear effect.
However, complete equilibration may in fact lead to long-range 
hypermagnetic fields~\cite{Figueroa:2017hun}.
These effects can be neglected as long as the growth rate
of the gauge fields is smaller than $ \gamma  $ in (\ref{kineqn}). 
In the Standard Model this requires that the chemical potential conjugate to
$ L _ {  e \rm R } $ satisfies (see appendix~\ref{a:ins}) 
\begin{align}
 | \mu  _ { \LeR } | \lsim  1.4 \cdot  10 ^{ -3 } T  
\label{constraint}
\end{align}
when $ \gamma  $ is comparable to the Hubble rate.

The processes contributing to the rate $ \gammaLL $ are very similar to those
in the production of ultrarelativistic sterile neutrinos
\cite{Besak:2012qm,Anisimov:2010gy}. 
There are two different types of contributions
at leading order, which is $ h _ e ^ 2 g ^ 2 $ where $ g $ denotes a
generic Standard Model coupling and $ h _ e $ is the electron Yukawa
coupling. The first type are $ 2 \to 2 $ scattering processes.
The second includes the (inverse) $ 1 \leftrightarrow 2 $ decay of
Higgs bosons into right-handed electrons and lepton doublets.  This
decay is kinematically allowed when the thermal
Higgs mass is sufficiently large.  One also has to take into account $
1n \leftrightarrow 2 n $ scatterings with soft gauge boson
exchanges. Due to their collinear nature  these  processes are not
suppressed.  On the contrary, they lead to strong enhancement compared
to the rate for Higgs decay, because they open several new channels,
which also happens in sterile neutrino
production~\cite{Anisimov:2010gy}.  Therefore the multiple scatterings
of $ 1n \leftrightarrow 2n $ particles with arbitrary $ n $ have to be
included, which is known as Landau-Pomeranchuk-Migdal (LPM)
resummation \cite{Landau:1953um,Landau:1953gr,Migdal:1956tc}.  A
complication compared to sterile neutrino production is that
right-handed electrons have Standard Model gauge interactions,
because they carry weak hypercharge. Therefore they are also affected
by multiple scattering, similar to gluons in QCD 
\cite{Baier:1996kr,Zakharov:1996fv,Arnold:2002ja}.

In section~\ref{s:thermal} we recall general expressions
for equilibration rates, and apply them to $ e _ { \rm R } $ equilibration.
In  section~\ref{s:lpm} we compute the contribution 
from (inverse) Higgs decays and $ 1n \leftrightarrow 2n $ 
processes. The $ 2\to 2 $ processes are treated in
section~\ref{s:2to2}.  Section~\ref{s:res} contains
numerical results and comparison with
previous work. Susceptibilities are computed in appendix~\ref{a:sus}.
The solution to the integral equation which
sums all $ 1n \leftrightarrow 2n $ 
processes is described in appendix~\ref{a:sol},
 and some integrals
for the $ 2 \to 2 $ processes are treated in appendix~\ref{a:I}.
We estimate the conversion of  $ L _ { e \rm R } $
into hypermagnetic fields through the chiral anomaly
and obtain the bound~\eqref{constraint} in appendix~\ref{a:ins}.

\paragraph{Notation and conventions} 
We write four-vectors in lower-case italics, $ k $,
and the corresponding three-vectors in boldface, $ \vec k $. Integrals
over three-momentum are denoted by $ \int _ { \vec k } \equiv
( 2 \pi ) ^ { -3 } \int d ^ 3 k  $. When working in imaginary time
we have four-vectors $ k = ( k ^ 0 , \vec k ) $ with $ k ^ 0 = \pi i n T $
with $ n $ even (odd) for bosons (fermions).
We denote fermionic Matsubara sums by a tilde,~%
$ \widetilde { \sum } _ { k ^ 0 } $.
We use the metric with signature $ (+,-,-,-) $.
Covariant derivatives are
  $ D _ \mu  =
  \partial _ \mu  + i y _ \alpha g ' B _ \mu  + \cdots $
with the hypercharge gauge coupling $ g ' $ and gauge field $ B $,
such that $ y _ 
  \varphi = 1/2 $ for the Higgs field $ \varphi  $.
The quartic term in the Higgs potential  is 
  $ \lambda ( \varphi ^\dagger \varphi ) ^ 2 $.

\section{Equilibration rates from thermal field theory}
\label{s:thermal}

\subsection{General considerations}
\label{s:master} 

We consider one or several 
charges $ Q _ a $ which are almost conserved, meaning that
they change much more slowly than most other (``fast'') 
degrees of freedom in the hot plasma.
The fast degrees of freedom equilibrate on a much
shorter time scale than the slow ones. 
We are interested in the long time behavior, that is on the time
evolution of the slow variables, so that the fast ones 
are always in equilibrium.
We assume small deviations $ \delta  Q _ a \equiv  Q _ a - Q _ a ^ {\rm eq } $
from thermal equilibrium, so that the time evolution of $ \delta  Q _ a $ is
determined by  linear equations.\footnote{%
In~\cite{Bodeker:2014hqa,Bodeker:2017deo} the 
conserved charges and the equilibrium values of 
the slow charges are assumed to vanish. Here we allow
for non-zero values for both of them.}
Without the Hubble expansion these effective kinetic equations 
are of the form
\begin{align}
  \dot Q _ a = - \gamma  _ { ab } \delta  Q _ b 
  \label{generalkin} 
   .
\end{align}
Both $ Q _ a ^ { \rm eq } $, and the rates
$ \gamma  _ { ab } $ only depend on the
temperature of the fast degrees of freedom, 
and of the values of the strictly conserved
charges. The rates 
can be written as \cite{Bodeker:2014hqa,Bodeker:2017deo} \footnote{%
We assume that the operators $ Q _ a $ commute at equal times.}
\begin{align}
  \gamma  _ { ab } =
  \frac 1 V
  \lim \limits _ { \omega \rightarrow 0 } 
\frac { 1 } {  \omega } { \rm Im } \Pi ^ { \rm ret } _ { ac } ( \omega )
\left( \Xi  ^ { - 1 } \right) _ { cb } 
\label{rate}
\end{align}
where $ V $ is the volume, which will be taken
to infinity, and 
\begin{align}
   \Pi  ^ { \rm ret } _ { ab } ( \omega  ) 
   \equiv 
   i
   \int \limits _ 0 ^ { \infty  } \! d t \, e ^{ i \omega  t } 
   \left \langle 
      \left [ \dot Q _ a ( t ) , \dot Q _ b ( 0 ) \right ]
   \right \rangle
   \label{Piret}
\end{align}  
is a retarded correlation function, which is computed in 
thermal equilibrium.
The matrix of susceptibilities $ \Xi  $ is determined by the
fluctuations of  $ \delta  Q _ a $ in thermal equilibrium,
\begin{align}
   \Xi  _ { ab }
   \equiv 
   \frac 1 { T V } \langle \delta  Q _ a \delta  Q _ b \rangle
   \label{Xi} 
   ,
\end{align}
taken at fixed 
values of strictly conserved charges.

At leading order in the $ Q _ a $-violating interaction strength 
one can
neglect these interactions in 
the expectation values in (\ref{Piret}) and (\ref{Xi}), so that
the $ Q _ a $ violation appears only in the operators 
$ \dot Q _ a $ in (\ref{Piret}).
Then  one can 
introduce chemical potentials $ \mu  _ a  $ for the 
$ Q _ a $.
For the charge densities
$ n _ a \equiv Q _ a /V $
we have 
at linear order 
\begin{align}
   \delta  n _ a = \Xi  _ { ab } \mu  _ b
   \label{deltana} 
   .
\end{align} 
Thus the kinetic  equations 
\eqref{generalkin} 
are equivalent to
\begin{align}
  \dot n _ a  = - \Gamma   _ { a b } \mu _ { b }
  \label{kineticmu}
  ,
\end{align}
with
\begin{align}
  \Gamma  _ { ab } 
  =
  \frac 1 V
  \lim \limits _ { \omega \rightarrow 0 } 
  \frac { 1 } {  \omega } { \rm Im } \Pi ^ { \rm ret } _ { ab } ( \omega )
  \label{Gamma}
   .
\end{align}
The coefficients $ \Gamma  _ { ab } $ are functions of the temperature
and of the values  of strictly conserved charges.

The equations (\ref{kineticmu}) for $  n _ a $ can be closed
as follows. One introduces chemical potentials not only for the
slowly varying charges but also for the strictly conserved ones. 
One computes the pressure $ P $ as a function of these chemical
potentials. Then the charge densities are given by 
\begin{align} 
    n _ A = \frac { \partial P } { \partial \mu  _ A } 
   , 
\end{align} 
where 
upper case indices label both slowly violated and
strictly conserved charges.
These are relations between all charge densities and
all chemical potentials. They can be solved to give
the chemical potentials of the slowly varying charges
in terms of all charge densities.

We assume all charge densities to vanish for zero chemical
potentials.
If the charge densities are sufficiently small,
the required relations are 
\begin{align} 
   n _ A = \chi  _ { AB } \mu  _ B
   \label{na} 
\end{align} 
with the matrix of susceptibilities
 \begin{align}
  \chi  _ { AB }
  \equiv 
  \left . \frac { \partial ^ 2 P } { \partial \mu  _ A \partial \mu  _ B }
  \right | _ { \mu  = 0 } 
  \label{chiPmu}
   .
\end{align}

\subsection{Right-handed electrons}
\label{s:rhe} 

Let us now apply the above formulas to the equilibration of 
right-handed electrons. 
In the Standard Model the lepton number 
\begin{align}
L _ { e { \rm R } }
\equiv 
\int d^3 x \,
    e _ { \rm R } ^ { \dagger } e _ { \rm R } ^ { \phantom \dagger } 
\label{L}
\end{align}
carried by right-handed electrons
is violated by their Yukawa interaction
\begin{align}
\mathscr L _ { { \rm int } } = 
- h _ e \overline { e _ { \rm R } } \varphi ^ \dagger \ell + { \rm H. c. }
\label{Lint}
\end{align}
with the Higgs field $ \varphi $ and the left-handed electron doublet 
$ \ell $. The electron Yukawa coupling $ h _ e $ is chosen to be real.
$ L _ { e { \rm R } } $ is also violated by the chiral anomaly 
which can lead to the creation of hypercharge 
magnetic fields \cite{Joyce:1997uy}.

The time derivative of $ L _ { e \rm R } $ due to the Yukawa 
interaction (\ref{L})  reads
\begin{align}
   \dot L _ { e { \rm R } } 
   = - i \int d ^ 3 x \left(h _ e 
    \overline { e _ { \rm R } } \varphi ^ \dagger \ell - {\rm H.c.}
    \right) 
   .
\label{dotZ}
\end{align}
This will be used 
to compute
\begin{align} 
   \GammaLL \equiv \Gamma  _ { \LeR \LeR } 
  .
\end{align} 
by means of (\ref{Piret}), (\ref{Gamma}).
The chiral anomaly term does not contribute to (\ref{Gamma}), since
the hypercharge gauge field is abelian, and 
unlike in non-abelian gauge theories the corresponding
winding number does not diffuse.

To determine the chemical potentials on the right-hand side
of~(\ref{Gamma}),  we have to 
identify the strictly conserved charges and determine their correlations. 
In the symmetric phase, the conservation of
baryon number $ B $ and the lepton numbers 
 $ L _ \alpha  $ in flavor $ \alpha  $ are 
violated by electroweak sphalerons. 
However, in the Standard Model the charges
\begin{align}
X _ \alpha \equiv L _ \alpha - \frac B 3
\label{X}
\end{align}
are conserved.
Furthermore, gauge charges are conserved.
In the symmetric phase the 
non-abelian gauge charges are not correlated with the weak hypercharge $ Y $ 
or with non-gauge charges. 
However, the correlation
of $ L _ { e \rm R } $ with 
the weak hypercharge does not vanish, $ \chi  _ { L _ { e \rm R } Y } 
\neq 0 $, and must be included in (\ref{na}).
In the imaginary time formalism 
the temporal component of the hypercharge gauge field
$ B _ 0 $ is purely imaginary. It has  
a non-zero expectation value \cite{Khlebnikov:1996vj},
which plays the role of a hypercharge
chemical potential,
\begin{align} 
   \mu  _ Y = i g' B _ 0 
   \label{muY} 
   ,
\end{align}
ensuring hypercharge neutrality.  

To illustrate the use of (\ref{kineticmu}) and (\ref{na})  
consider three important examples. 
We need the inverse of the 
susceptibility matrix $ \chi  $, which is computed 
in appendix~\ref{a:sus}.
\begin{enumerate}
\item 
Only Standard Model interactions, $ L _ { e { \rm R } } $
is the only slow variable. Assume $ X _ e $ to be non-zero,
and $ X _ \mu = X _ \tau  = 0 $. Then (\ref{na})  gives
\begin{align} 
   \label{mun}
    \mu  _ {  L _ { e { \rm R } } }  
   =
   \left ( \chi  ^{ -1 } \right ) 
   _ { L _ { e { \rm R } }  L _ { e { \rm R } } } 
   n _ { L _ { e  \rm R } }
   + 
   \left ( \chi  ^{ -1 } \right ) 
   _ { L _ { e { \rm R } }  X _ { e  } } 
   n _ { X _ e }
   .
\end{align}
which we write as
\begin{align} 
   \label{muneq}
   \mu  _ { \LeR } 
   =
   \left ( \chi  ^{ -1 } \right ) _ { \LeR \LeR } 
   \left ( n _ { \LeR } - n _ { \LeR } ^{ { \rm eq } } \right ) 
\end{align} 
with
\begin{align} 
   \label{neq}
   n _ { \LeR } ^ { \rm eq } 
   =
   - \frac { \left ( \chi  ^{ -1 } \right ) _ { \LeR X _ e }
      }
   {\left ( \chi  ^{ -1 } \right ) _ { \LeR \LeR } }
   \, n _ { X _ e } 
\end{align} 
Combining this with (\ref{kineticmu}) yields
\begin{align}
   \dot  n _ { L _ { e { \rm R } } } 
  =
   -\gammaLL
   \left ( n _ { L _ { e { \rm R } } } 
   -  n _ { \LeR   } ^ { \rm eq } \right )
   \label{dotnnneq}
\end{align} 
with 
\begin{align}
   \gammaLL = \left ( \chi  ^{ -1 } \right ) _ { \LeR \LeR } 
   \GammaLL
   \label{gchi}
   .
\end{align}  
Equation (\ref{chiinveR}) then gives 
\begin{align} 
   \gammaLL
   &= 
   \frac { 4266 } { 481 } 
   T ^{ -2 } 
   \GammaLL 
   \label{gGexplicit}
   ,
\\
     n _ { L _ { e { \rm R } } } ^ { \rm eq }  
   &=
   \frac { 185 } { 711 } n _ { X _ e } 
   .
   \label{neqex} 
\end{align} 

\item
Only Standard Model interactions, $ L _ { e { \rm R } } $
is the only slow variable. Now allow for all $ X _ \alpha $
to be non-vanishing, with the constraint $ B - L = \sum _ { \alpha } X _ \alpha = 0 $.
Then we have  
\begin{align}
   \dot  n _ { L _ { e { \rm R } } } 
  = - 
   \GammaLL
   \left [ 
       \left ( \chi  ^{ -1 } \right ) 
      _ { L _ { e { \rm R } }  L _ { e { \rm R }   } } 
   n _ { L _ { e { \rm R } } }
   + \sum _ \alpha  
       \left ( \chi  ^{ -1 } \right ) 
      _ { L _ { e { \rm R } }  X _ { \alpha } } n _ { X _ \alpha } 
   \right ] 
   \label{ndotlg} 
   .
\end{align} 
By means of (\ref{chiinveR}) 
this turns into
\begin{align}
   \dot  n _ { L _ { e { \rm R } } } 
  = - 
   T ^{ -2 } \GammaLL
   \left [ 
      \frac { 4266 } { 481 } n _ { L _ { e { \rm R } } }
      - \frac { 30 } { 13 }  n _ { X _ e } 
      + \frac { 24 } { 37 } \left (  n _ { X _ \mu } + n _ { X _ \tau  } 
          \right )
  \right ] 
   \label{lg} 
   .
\end{align}
This equation can be 
recast in the form of~\eqref{dotnnneq} with~\eqref{gGexplicit}
and 
\begin{align}
    n _ { L _ { e { \rm R } } } ^ { \rm eq }  
   =
   \frac { 1 } { 3 } n _ { X _ e } 
   \label{ex2} 
   ,
\end{align}
which agrees with the result in \cite{Cline:1993bd}.

\item
Type-I see-saw models realizing leptogenesis. 
Here one supplements the Standard Model with right-handed Majorana neutrinos
whose Yukawa interactions violate $ X _ \alpha $-conservation.
If leptogenesis takes place around the same time as the equilibration of right-handed
electrons,  then  
both the $ X _ \alpha $ and $ L _ { e {\rm R } } $ have to be treated
as slow variables. 
The time derivatives of $ X _ \alpha $ are uncorrelated
with (\ref{dotZ}), and therefore the rate coefficients 
$ \Gamma  _ { L _ { e \rm R } X _ \alpha } $ vanish.
In this case~\eqref{ndotlg} and~\eqref{lg} hold again, and so does
\eqref{gGexplicit}. This time the terms with $ n _ { X _ { \alpha } } $ 
do not contribute to $ n _ { L _ { e { \rm R } } } ^ { \rm eq } $, 
but constitute
individual source terms, so that $ n _ { L _ { e \rm R } } ^ { \rm eq } = 0 $
and 
\begin{align}
   \gamma  _ { L _ { e \rm R } X _ e }
   &= 
   -
   \frac { 30 } { 13 } 
   T ^{ -2 } 
   \GammaLL
   \label{gLXe} 
   ,
\\
   \gamma  _ { L _ { e \rm R } X _ \alpha }
   &= 
   \frac { 24 } { 37 } 
   T ^{ -2 } 
   \GammaLL
   \qquad ( \alpha = \mu, \tau ) 
   \label{gLXa} 
   .
\end{align}

\end{enumerate} 

We evaluate 
$ \GammaLL $  at vanishing chemical
potentials, which is appropriate when
the charge densities are small. 
This way we avoid the problem of infrared divergences in processes
with Higgs bosons in the initial or 
final state (see~\cite{Ghiglieri:2017gjz}).

\section{Higgs decay and multiple soft scattering
}
\label{s:lpm}

The bulk of particles in the plasma have `hard' momenta, $ p \sim T $.
In the symmetric phase, the Standard Model 
particles carry thermal  masses. For the Higgs boson the
thermal mass is momentum independent and is given by
\cite{Weldon:1982bn} 
\begin{align}
  m _ { \varphi } ^ 2
  &= \frac { 1 } { 16 }
  \left[
    3 g ^ 2 + { g ' } ^ 2
    + 4 h _ t ^ 2 + 8 \lambda \right] ( T ^ 2 - T _ 0 ^ 2 ) 
   ,
   \label{m2higgs}
\end{align}
with $ T _ 0 = 160 $ GeV.
For hard fermions one has to use the so-called asymptotic
thermal masses \cite{Weldon:1982bn}, which for the left- and right-handed
leptons are given by
\footnote{%
For fermions 
the asymptotic mass is a factor $ \sqrt 2 $ larger 
than the one at zero momentum \cite{Weldon:1982bn}.
In \cite{Cline:1993bd} the zero-momentum fermion masses are used.}
\begin{align} 
m _ { \ell } ^ 2 &= \frac { 1 } { 16 } \left[
                       3 g ^ 2 + { g ' } ^ 2 \right]  T ^ 2   ,
   \label{m2ell}\\
m _ { e _ { \rm R } } ^ 2 &= \frac { 1 } { 4 } { g ' } ^ 2 T ^ 2
   . 
   \label{m2eR}
\end{align}
\begin{figure}[t]
 \centering
 \includegraphics[width=.3\textwidth]{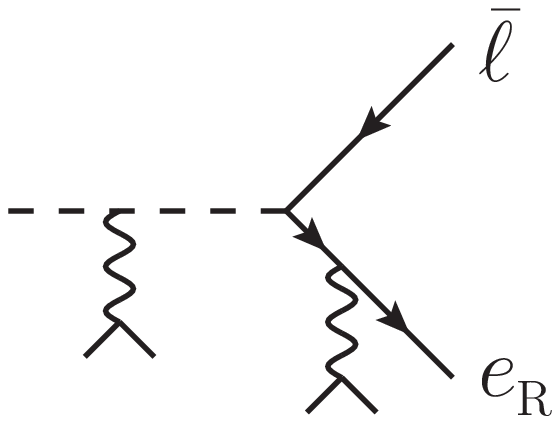}
 \hspace{.14\textwidth}
 \includegraphics[width=.3\textwidth]{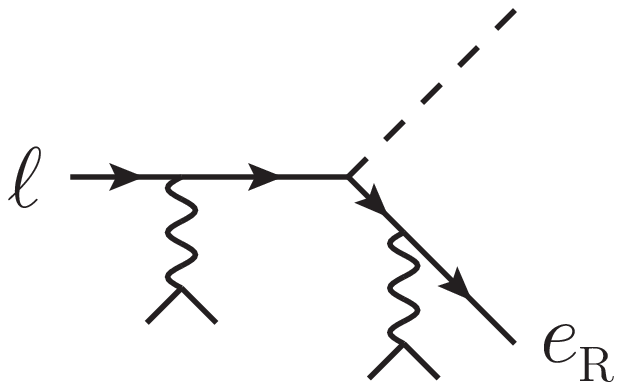}
 \caption{The interference of these two exemplary $ 1n\to 2n $ processes with $ n = 2 $
         needs to be taken into account. The gauge bosons have soft momenta
         $ q \sim gT $.}
 \label{f:1n2n}
 \end{figure}%
For $ T \gg T _ 0 $  the 
Higgs bosons have the largest mass, and for certain values
of the couplings their decay into 
left-handed lepton doublets $ \ell $ and
the right-handed electrons is kinematically allowed. 
With increasing temperature the top Yukawa coupling decreases 
such that above a certain temperature $ m _ \varphi $ becomes
smaller than $ m _ \ell 
+ m _ { e _ { \rm R } } $
and the channel closes. Since 
$ m _ \varphi > m _ \ell  > m _ { e _ { \rm R } } $
at any temperature well above the electroweak scale,
no other $ 1 \leftrightarrow 2 $ decay channel
opens up at a higher temperature.

Since the masses are small compared to $ T $,   the particles
participating in the decay process
are 
ultrarelativistic. Furthermore, their momenta 
are nearly collinear,
with transverse momenta $ p _ \perp $ of order $ g T $. The wave packets of
the decay products have a width of order $ 1/p _ \perp $. 
They overlap for  a time of order $ 1/(g ^ 2 T ) $, the so-called
formation time. Here the formation time is of the same order of magnitude
as the mean free time between scatterings with `soft' momentum transfer
$ q \sim g T $.
Thus the particles typically scatter multiple times before their wave packets
separate, so that the scatterings cannot be treated independently.
We show two exemplary diagrams in figure~\ref{f:1n2n}.
This situation is similar to bremsstrahlung in a medium
in QED \cite{Landau:1953um,Landau:1953gr,Migdal:1956tc,Arnold:2001ba} (see
also \cite{Aurenche:2002wq}), 
and in QCD 
\cite{Baier:1996kr,Zakharov:1996fv,Baier:1996sk,Zakharov:1997uu,Arnold:2002ja}, 
where it leads
to a suppression of the emission probability.
In the case of sterile neutrino production, on the other hand, 
it gives a strong enhancement, because new kinematic channels are opened
\cite{Anisimov:2010gy}.
We compute the Higgs decay  in
section~\ref{s:tree}, and include
multiple soft scatterings in section~\ref{s:multiple}.

\subsection{Higgs decay}
\label{s:tree}

We start from the imaginary-time correlator
\begin{align}
\Pi _ { L _ { e { \rm R } } L _ { e { \rm R } } } ( i \omega _ n )
 =
\int \limits _ 0 ^ { 1/T }  \! d \tau \, e ^ { i \omega _ n \tau } 
   \langle \dot L _ { e { \rm R } }  ( - i \tau )
\dot L _ { e \rm R }  ( 0 ) \rangle,
\label{mats}
\end{align}
with bosonic frequency $ \omega _ n $.
Without soft gauge interactions, (\ref{mats}) reads
\begin{align}
   \Pi _ { L _ { e { \rm R } } L _ { e { \rm R } } } ( i \omega _ n )
 =
 -
2 V h _ e  ^ 2 T ^ 2 \widetilde { \sum \limits _ { p ^ 0, k ^ 0 } } \int _ { \vec p, \vec k }
{ \rm tr } \left[ S _ { \ell } ( p ) S _ { e _ { \rm R } } ( k ) \right] 
\Delta _ { \varphi } ( p - k + i \omega _ n  u)
  \nonumber  \\
+ ( i \omega _ n \rightarrow - i \omega _ n ),
\label{matstree}
\end{align}
with $ u = ( 1, \vec 0 ) $ the four-velocity of the plasma.
We write the scalar field propagator as
\begin{align} 
\Delta _ a ( p ) 
   &= \frac { - 1 } 
    { ( v \cdot p ) ( \overline v \cdot p ) - \vec p _ \perp ^ 2 - 
   m _ a ^ 2 }
    \label{scalarprop}
    .
\end{align}
Here, $ v = (1, \vec v) $ with a unit vector
$ \vec v $, which defines the longitudinal direction, and
$ \overline v = (1, - \vec v ) $.
Chiral symmetry is unbroken, even with thermal masses. Therefore
the non-vanishing components of the fermion propagators
in the Weyl representation can be written as 2$ \times $2 matrices,
\begin{align}
S _ { \ell } ( p ) 
   &= 
   \sigma \cdot p \Delta  _ \ell ( p ) 
    \label{ellprop} 
    , \\
S _ { e _ { \rm R } } ( p ) 
   &= 
   \overline \sigma \cdot p \Delta  _ { e _ { \rm R } } ( p ) 
    \label{eRprop} 
   ,
   \end{align}
where $ \sigma ^ \mu $, $ \overline \sigma ^ \mu  $ 
are  the usual Pauli matrices.  
There are two different kinematic situations which we have
to take into account: either 
all momenta satisfy $ v \cdot p \sim g ^ 2 T $,  
$ \overline v \cdot p \sim T $
or the same but with $ v \leftrightarrow \overline v $.
The second case gives the same result as the first but with $ i \omega  _ 
n \to - i \omega  _ n $.
For $ v \cdot p \sim g ^ 2 T $ the scalar propagator 
can be approximated as
\begin{align} 
   \Delta  _ a ( p ) = \frac 1 { 2 p _ \parallel } 
    D _ a  ( p ) 
    \label{Delta}
\end{align} 
where $ p  _ \parallel \equiv \vec v \cdot \vec p $ is the large component
of $ \vec p $, and 
\begin{align} 
   D _ a ( p ) \equiv \frac { -1 } { v \cdot p - ( \vec p _ \perp ^ 2 
   + m _ a ^ 2 ) /( 2 p _ \parallel ) }
   \label{Da} 
   .
\end{align} 
Similarly, the fermion propagators can be written 
as (see \emph{e.g.}\ \cite{Anisimov:2010gy})
\begin{align}
S _ { \ell } ( p )
   &=
    \eta ( \vec p ) \eta ^ \dagger ( \vec p ) \, D _ { \ell } ( p )
\label{Sell}
,
\\
S _ { e _ { \rm R } } ( p )
&=
    \chi ( \vec p ) \chi ^ \dagger ( \vec p ) \, D _ { e  _ { \rm R } } ( p )
\label{SeR}
\end{align}
with the spinors 
\begin{align}
\eta (\vec  p ) 
&= \left[ 1
  - \frac { 1 } { 2 p _ \parallel } \left( {\bm \sigma} \cdot \vec p _ \perp \right)
  \right] \begin{pmatrix} 0 \\ 1 \end{pmatrix}
  \label{spinoreta}
  ,
  \\
\chi ( \vec p ) 
  	&= 
	\left[ 1
  + \frac { 1 } { 2 p _ \parallel } 
	\left( {\bm \sigma} \cdot \vec p _ \perp \right)
  \right] \begin{pmatrix} 1 \\ 0 \end{pmatrix}
	.
\label{spinorchi}
\end{align}
In~\eqref{Sell} through~\eqref{spinorchi} we keep only the 
leading order contributions to the equilibration rate. 
It is convenient to associate the spinors in (\ref{Sell}), (\ref{SeR}) 
with the adjacent vertices rather than with the propagators.

After performing the sum over Matsubara
frequencies we encounter a factor 
\begin{align}
   \mathscr F ( p _ \parallel, k _ \parallel ) 
   = 
   f _ { \rm F } ' ( k _ \parallel ) 
   \left[ f _ { \rm F } ( p _ \parallel ) 
   + f _ { \rm B } ( p _ \parallel - k _ \parallel ) \right]
   .
\label{F}
\end{align}
We can then analytically continue $ i \omega _ n $ to arbitrary complex
$ \omega   $ which gives 
\begin{align}
    \Pi  _ { L _ { e { \rm R } } L _ { e \rm R }  } ( \omega ) 
   =
   2 h _ e ^ 2 V 
   \int _ { \vec k, \vec p } 
   \frac { \mathscr F ( p _ \parallel , k _ \parallel ) }
                                    {  k _ \parallel - p _ \parallel  }
   \, \eta ^ { \dagger } ( \vec p )
   \chi ( \vec k )
    \chi  ^\dagger ( \vec k ) \eta  ( \vec p ) 
   \frac { \delta  E } { \delta  E  - \omega  } 
{} +  ( \omega   \to - \omega     ) 
\label{conti}
   ,
\end{align}
where
\begin{align} 
   \delta  E 
   = 
   \delta E ( \vec p _ \perp, \vec k _ \perp )
 \equiv 
 \frac {  m _ { e _ { \rm R } } ^ 2  + \vec k _ \perp ^ 2 } { 2 k _ \parallel }
- \frac { m _ { \ell } ^ 2 + \vec p _ \perp ^ 2 } { 2 p _ \parallel }
- \frac { m _ { \varphi } ^ 2 
   + (\vec k _  \perp - \vec p _ \perp ) ^ 2 } 
   { 2 (k _ \parallel - p _ \parallel )}
   \label{deltaE}
\end{align}
is the change of energy in the decay $ \varphi  \to \ell \, \overline 
e _ { \rm R } $.
When we take the imaginary part
of the retarded correlator 
\begin{align}
    \Pi ^ { \rm ret } _ { L _ { e { \rm R } } L _ { e \rm R }  } ( \omega ) 
= \Pi _ { L _ { e { \rm R } } L _ { e \rm R }  } ( \omega + i 0 ^ + ) 
   \label {ret}
\end{align} 
with $ \omega $ real,
  $ \delta  E $
becomes equal to $ \pm \omega $. For both signs one obtains the
same imaginary part.
Since we need this to compute the rate using~(\ref{rate}) we can drop
terms of order $ \omega ^ 2 $  and higher.
Pulling out
a factor $ \eta  ^\dagger ( \vec p ) \chi  ( \vec k ) $, corresponding
to the leftmost vertex in 
figure~\ref{f:multiple} (without gauge bosons) we may write
\begin{align}
    { \rm Im } 
   \Pi ^ { \rm ret }   _ { L _ { e { \rm R } } L _ { e \rm R }  } ( \omega ) 
   =   
   8 h _ e ^ 2 V 
   \omega  
   { \rm Im } 
   \int _ { \vec k, \vec p } 
   \frac { \mathscr F ( p _ \parallel , k _ \parallel ) }
                                    {  k _ \parallel - p _ \parallel  }
   \, \eta ^ { \dagger } ( \vec p ) 
   \chi ( \vec k ) \, j ( \vec p _ \perp, \vec k _ \perp )
\label{LPMtpf}
\end{align} 
where $ j $ satisfies 
\begin{align}
   \big( \delta  E - i 0 ^ + \big)
   j ( \vec p _ \perp, \vec k _ \perp ) 
   = 
   \frac 12 \chi  ^\dagger ( \vec k ) \eta ( \vec p )
   \label{jdecay}
   .
\end{align}  
Note that in the integrand of (\ref{LPMtpf}) the delta function 
$ \delta  ( \delta  E ) $ appears which enforces energy conservation for
the (inverse) Higgs decay. The coefficient $ \GammaLL $ is then obtained
by plugging (\ref{LPMtpf}) into (\ref{Gamma}).

\subsection{Multiple soft gauge boson scattering}
\label{s:multiple} 

 \begin{figure}
 \centering
 \includegraphics[width=0.6\textwidth]{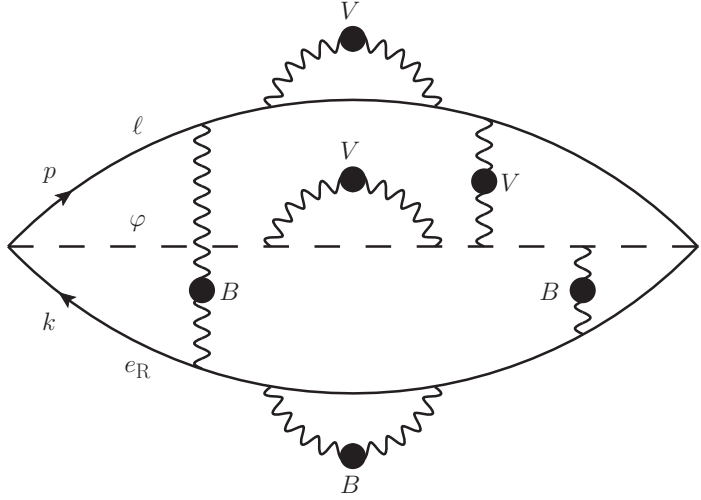}
 \caption{Typical diagram of multiple soft scattering the imaginary 
          part of which gives a contribution to the equilibration rate.
          $ V $ stands for either a $ W $ or a hypercharge gauge boson $ B $.
          All gauge bosons are soft and their propagators are hard
          thermal loop (HTL) resummed, as indicated by the thick dots.}
 \label{f:multiple}
 \end{figure}
 
Now we include the effect of multiple scattering mediated 
by soft gauge bosons, as sketched in figure~\ref{f:multiple}.\footnote{%
The range of the gauge interactions is one power of $ g $ smaller than
 the mean free path of the
fermions and the Higgs.
Therefore crossed gauge bosons, rainbow self-energies, or gauge boson
vertex corrections do not contribute at leading order
.}
The result can again be described by (\ref{LPMtpf}), where $ j $ now
satisfies
\begin{align}
\delta E ( \vec p _ \perp, \vec k _ \perp )  
   \; 
   j ( \vec p _ \perp, \vec k _ \perp )
   \nonumber 
     &  
   - 
   i \int \frac { d ^ 2 q _ \perp } { ( 2 \pi ) ^ 2 } 
   \Big\{ \mathscr C ( \vec q _ \perp ^ 2 )
   \big[ 
         j ( \vec p _ \perp, \vec k _ \perp ) 
      - j ( \vec p _ \perp - \vec q _ \perp, \vec k _ \perp ) 
   \big]\\
   { }     + \mathscr C ' ( \vec q _ \perp ^ 2 )
       \Big( 
     y _ \varphi    y _ \ell  
   \big [ 
     & 
      j ( \vec p _ \perp, \vec k _ \perp ) 
      - j ( \vec p _ \perp - \vec q _ \perp, \vec k _ \perp ) 
   \big ] 
   \nonumber \\
    { } + y _ \ell \, y _ { e _ { \rm R } } 
   \big [ 
   &
   j ( \vec p _ \perp, \vec k _ \perp ) 
   - j ( \vec p _ \perp - \vec q _ \perp, \vec k _ \perp - \vec q _ \perp) 
   \big ] 
   \nonumber 
   \\
      { }  - y _ \varphi \, y _ { e _ { \rm R } } 
    \big [ 
   &
   j ( \vec p _ \perp, \vec k _ \perp ) 
      - j ( \vec p _ \perp, \vec k _ \perp - \vec q _ \perp ) 
    \big ] 
       	    \Big) \Big\} 
   =
  \frac 12 \chi ^ { \dagger } ( \vec k )   \eta ( \vec p )
   .
\label{intequ}
\end{align}
Here $ \vec q _ \perp $ is the transverse momentum of 
an  exchanged gauge boson.  We have introduced 
\begin{align}
\mathscr C ( \vec q _ \perp ^ 2 )  
   & \equiv 
  \frac 3 4 g ^ 2 T
  \left( \frac { 1 } { \vec q _ \perp ^ 2 } 
   -  \frac { 1 } { \vec q _ \perp ^ 2 + m _{\rm D} ^ 2 } \right) 
   , 
   \label{C}
   \\
\mathscr C ' ( \vec q _ \perp ^ 2 ) 
   & 
   \equiv  { g ' } ^ 2 T
  \left( \frac { 1 } { \vec q _ \perp ^ 2 } 
   -  \frac { 1 } { \vec q _ \perp ^ 2 + { m _ {\rm D} ' } ^ 2}\right)
   \label{Cprime}
\end{align}
with the Debye masses \cite{Carrington:1991hz}
\begin{align}
m _ {\rm D} ^ 2 = \frac { 11 } { 6 } g ^ 2 T ^ 2,
\quad\quad
{ m _ {\rm D} ' } ^ 2 = \frac { 11 } { 6 } { g ' } ^ 2 T ^ 2
\label{Debye}
.
\end{align}
In the integral in~\eqref{intequ} the terms 
containing $ j ( \vec p _ \perp, \vec k _ \perp ) $ correspond to
self-energy insertions, which can be 
easily checked by an explicit calculation.%
\footnote{Note that
 $ y _ \varphi y _ \ell + y _ \ell y _ { e _ {\rm R } } - y _ \varphi y _ { e _ { \rm R } } 
   =
   \left( y _ \varphi ^ 2 + y _ \ell ^ 2 + y _ { e _ { \rm R } } ^ 2 \right) / 2 $.
 } The terms  with  $ \mathscr C $ and  $ \mathscr C '$ correspond to 
interactions mediated by
$ W $ or 
$ B $ bosons, respectively.
By themselves, the self-energies are infrared divergent due to the 
$ 1/\vec q _ \perp ^ 2$ term in $ \mathscr C $ and $ \mathscr C '$.
The subtracted terms in the square brackets in (\ref{intequ}) 
correspond to gauge boson exchange between different particles and
render the $ \vec q _ \perp $-integrals finite. 
The first two square brackets in (\ref{intequ}) 
also appear in the computation of the production rate of
ultrarelativistic sterile
neutrinos~\cite{Anisimov:2010gy}. The other two  
represent the exchange of weak hypercharge
gauge bosons by the  
right-handed electrons.
Replacing the integral in (\ref{intequ}) by 
$ 0 ^ + $, 
one neglects multiple soft scatterings and one recovers the equation 
(\ref{jdecay}) 
describing Higgs decay. 

Thanks to
three-dimensional rotational invariance,
the  solution to (\ref{intequ}) can be found as a function of a single 
transverse momentum \cite{Arnold:2015qya},
\begin{align}
j ( \vec p _ \perp, \vec k _ \perp )
=
J ( \vec P ),
\label{ansatz}
\end{align}
with
\begin{align}
\vec P \equiv x _ k \vec p _ \perp &- x _ p \vec k _ \perp
\label{P}
   ,
\\
x _ k \equiv \frac { k _ \parallel } {  p _ \parallel - k _ \parallel },
\quad & \quad
x _ p \equiv \frac { p _ \parallel } { p _ \parallel - k _ \parallel  }
   .
\label{xkxp}
\end{align}
In fact,~\eqref{deltaE} now takes the simple form 
\begin{align}
\delta E  
 =
\beta \left( \vec P ^ 2 + M ^ 2 \right)
\label{deltaEP}
\end{align}
with
\begin{align}
\beta &\equiv \frac { p _ \parallel - k _ \parallel } { 2 p _ \parallel k _ \parallel }
\label{beta}
,
\end{align}
and 
\begin{align} 
M ^ 2
&\equiv
\beta ^ { - 1 }
\left[
\frac {  m _ { e _ { \rm R } } ^ 2 } { 2 k _ \parallel }
- \frac { m _ { \ell } ^ 2 } { 2 p _ \parallel }
- \frac { m _ { \varphi } ^ 2 } { 2 (k _ \parallel - p _ \parallel )}
\right]
\label{M2}
   .
\end{align}
The right-hand side of (\ref{intequ}) turns into
\begin{align} 
  \frac 12 \chi ^ { \dagger } ( \vec k )   \eta ( \vec p )
  =
  - \frac \beta 2
	\left ( P _ x - i P _ y \right ) 
   \label{cde}
   .
\end{align} 
The function $ J ( \vec P ) $ can be expressed as  
\begin{align}
J ( \vec P ) 
   = 
   \frac { i \beta } { 4 } \left[ f _ x(\vec P) - i f _ y(\vec P) \right]
   ,
\label{jf}
\end{align}
where the two-component vector $ \vec f $ is a 
solution to 
\begin{align}
- i
\delta E 
   \; 
   \vec f ( \vec P )
   \nonumber   
   - \int \frac { d ^ 2 q _ \perp } { ( 2 \pi ) ^ 2 } 
   \Big\{ \mathscr C ( \vec q _ \perp ^ 2 )
   \big[ & 
         \vec f ( \vec P ) 
      - \vec f ( \vec P - x_k \vec q _ \perp )
   \big]\\
   { }     + \mathscr C ' ( \vec q _ \perp ^ 2 )
       \Big( 
     y _ \varphi    y _ \ell  
   \big [ 
     & 
         \vec f ( \vec P ) 
      - \vec f ( \vec P - x_k \vec q _ \perp )
   \big ] 
   \nonumber \\
    { } + y _ \ell \, y _ { e _ { \rm R } } 
   \big [ 
   &
         \vec f ( \vec P ) 
      - \vec f ( \vec P + \vec q _ \perp )
   \big ] 
   \nonumber 
   \\
      { }  - y _ \varphi \, y _ { e _ { \rm R } } 
    \big [ 
   &
         \vec f ( \vec P ) 
      - \vec f ( \vec P + x_p \vec q _ \perp )
    \big ] 
       	    \Big) \Big\} 
   =
  2 \vec P
\label{intf}
.
\end{align}
This is  the same integral equation as in 
\cite{Anisimov:2010gy} (with the appropriate hypercharge
assignments), but with two additional terms
representing the gauge interaction of right-handed electrons.

Now we choose the unit vector $ \vec v $ in the direction of  $ \vec k $.
Using 
$ \vec f ( \vec P ) \propto \vec P $ and integrating over 
the transverse momentum
$ \vec k _ \perp $, we obtain
\begin{align}
\GammaLL ^ { 
   \rm LPM } 
=
\frac { h _ e  ^ 2 }
      { 8 \pi ^ 3 }
\int\limits _ 0 ^ \infty \!    dk \!
\int\limits _ { - \infty } ^ \infty \! \! d p _ \parallel \,
\frac { \left( p _ \parallel - k \right) ^ 3 } { p _ \parallel ^ 2 k ^ 2 }
\mathscr F ( p _ \parallel, k )
\;
{ \rm Re } \!
\int \frac { d ^ 2 P } { ( 2 \pi ) ^ 2 }
\vec P \cdot \vec f ( \vec P )
\label{LPMrateP}
\end{align}
for the rate coefficient.
We solve~\eqref{intf} using the algorithm 
described in \cite{Anisimov:2010gy} and numerically
integrate~\eqref{LPMrateP} (see appendix~\ref{a:sol}).

 \begin{figure}[t]
 \includegraphics[width=.32\textwidth]{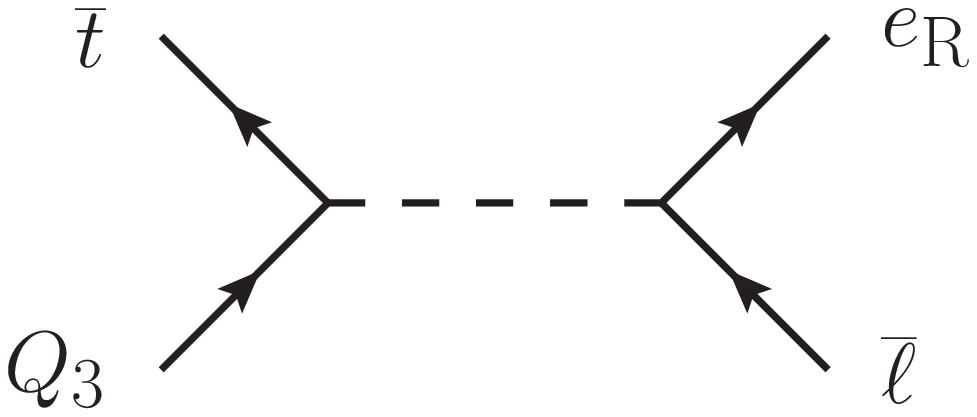}
 \includegraphics[width=.32\textwidth]{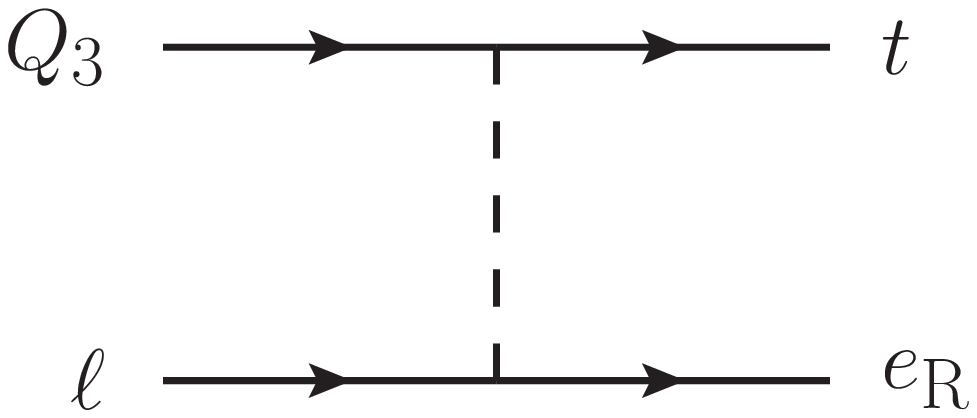}
 \includegraphics[width=.32\textwidth]{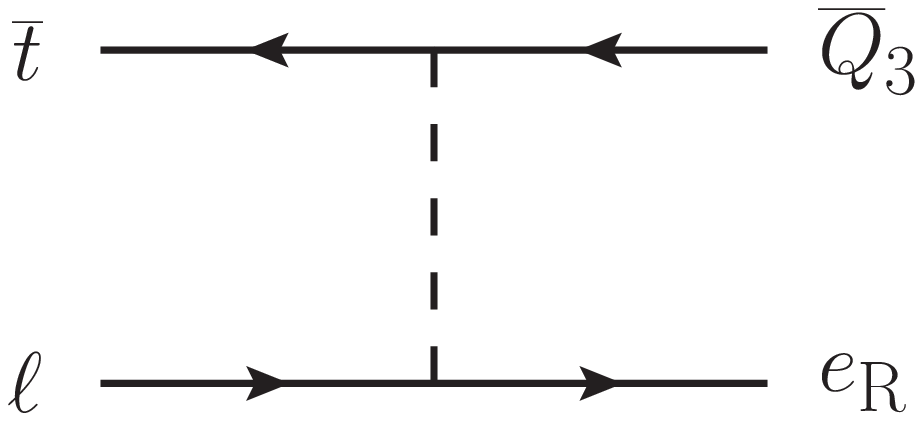}
 \\[1cm]
 \includegraphics[width=.32\textwidth]{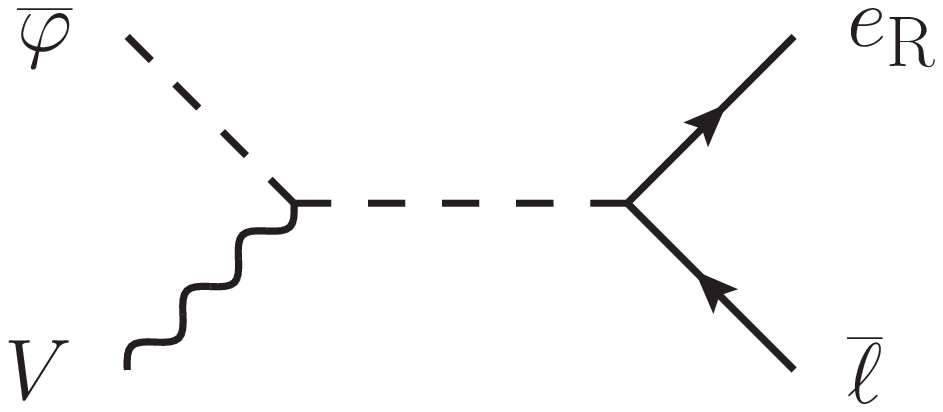}
 \includegraphics[width=.32\textwidth]{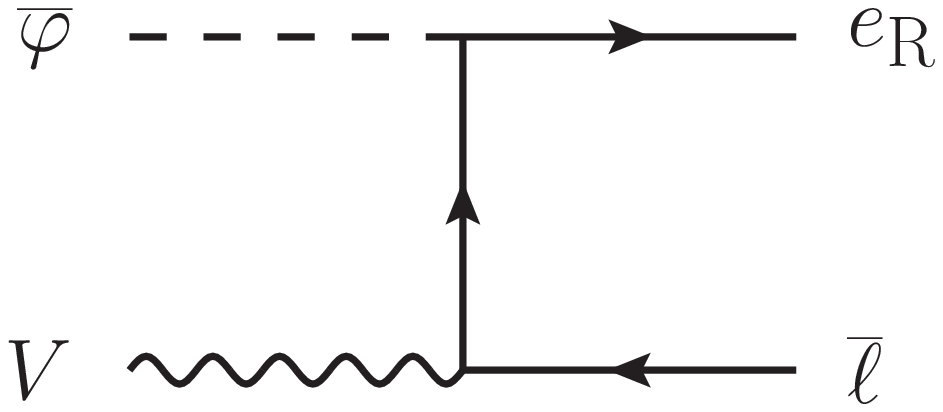}
 \includegraphics[width=.32\textwidth]{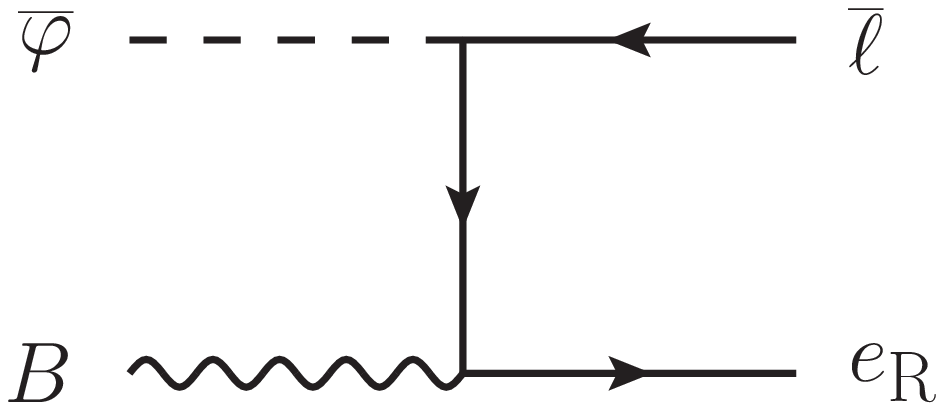}
 \\[1cm]
 \includegraphics[width=.32\textwidth]{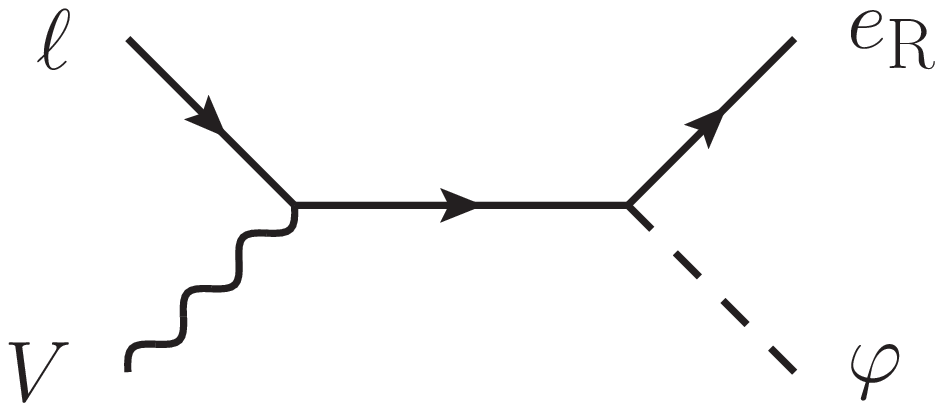}
 \includegraphics[width=.32\textwidth]{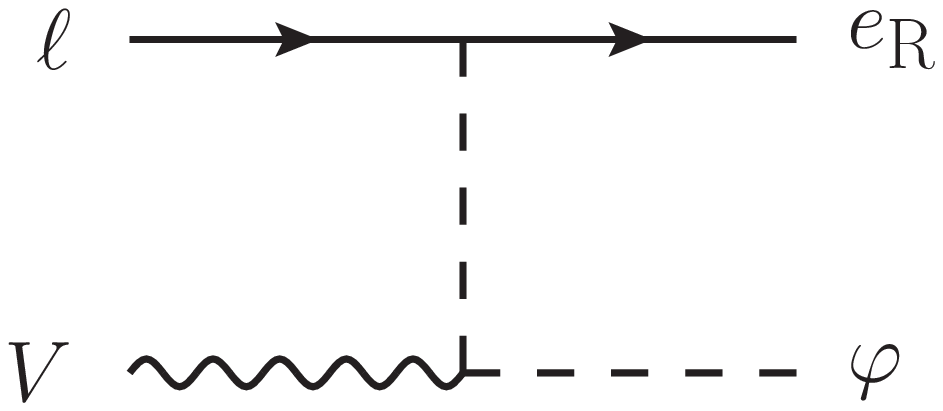}
 \includegraphics[width=.32\textwidth]{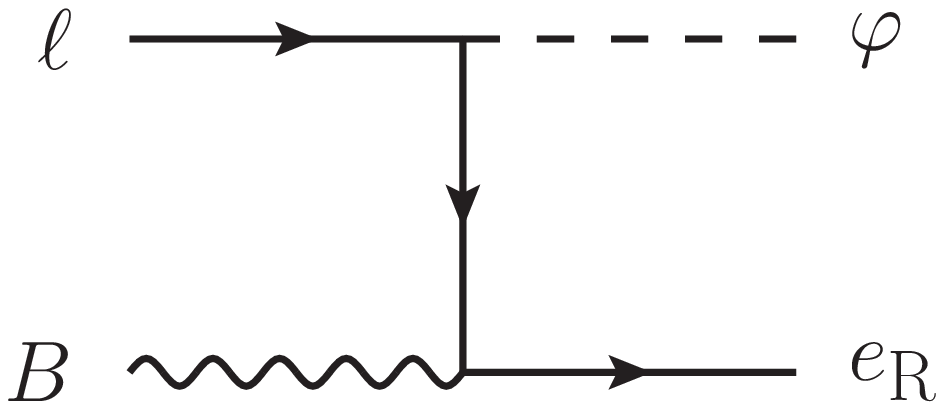}
 \\[1cm]
 \includegraphics[width=.32\textwidth]{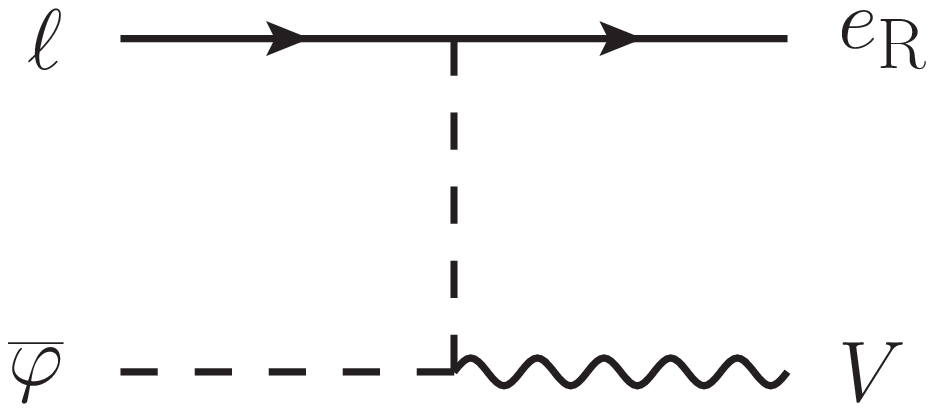}
 \includegraphics[width=.32\textwidth]{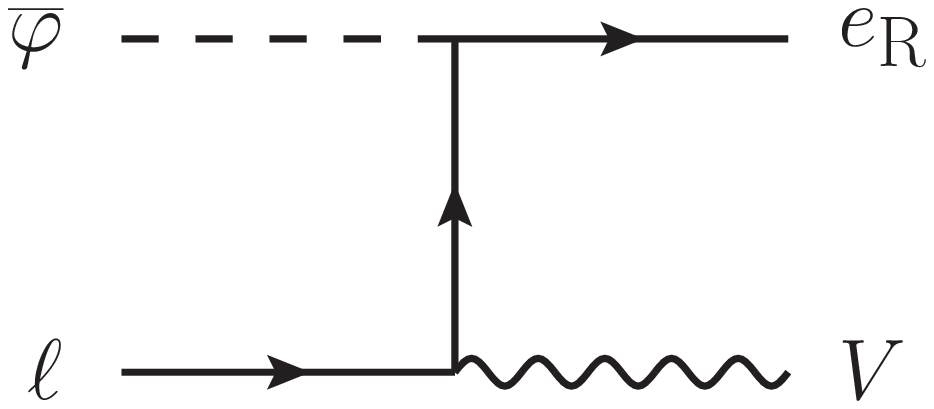}
 \includegraphics[width=.32\textwidth]{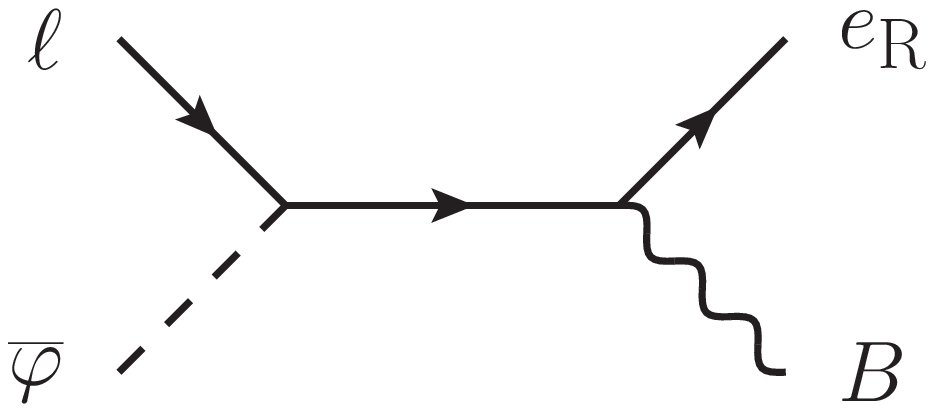}
 \caption{Diagrams for the $ 2 \rightarrow 2 $ processes.
 First line: Quark contributions,
 second line:
 $ V \overline \varphi \rightarrow \overline \ell e _ { \rm R } $,
 third line:
 $ V \ell \rightarrow \varphi e _ { \rm R } $,
 fourth line:
 $ \ell \overline \varphi \rightarrow V e _ { \rm R }$.
 Here and in the diagrams we denote $ V = B, W $.
 The exchanged fermion in $ t $-channel is an $ \ell $
 in the second column and an $ e _ { \rm R } $ in the
 third column.
 }
 \label{f:2to2diag}
 \end{figure}

\section{$ 2 \rightarrow 2 $ processes}
\label{s:2to2}

At order $ h _ e ^ 2 g ^ 2 $ there are also contributions from $ 2 \rightarrow 2 $
scatterings. The corresponding diagrams are shown in figure~\ref{f:2to2diag}.
At leading order all external particles have 
hard momenta, $ p \sim T $,
and one can neglect
their thermal masses. 
For $ s $-channel exchange the internal momenta are hard as well, and one
can neglect thermal effects on the propagators.\footnote{%
In \cite{Cline:1993bd} the thermal Higgs mass is included in
the Higgs propagator for the process $ \overline t Q _ 3 \to \ell e 
_ { \rm R } $. This leads to the complication that the propagator
can become on-shell, and  a subtraction has to be performed. This
problem does not arise in a strict leading order calculation.} 
However, 
momenta exchanged in the $ t $-channel become soft at leading order. 
We treat these
contributions in section~\ref{s:soft}.

Again, the processes are similar to the ones
encountered in relativistic sterile neutrino 
production in \cite{Besak:2012qm}. 
However, as in the case of the $ 1n \leftrightarrow 2n $ processes, 
in $ e _ { \rm R } $ equilibration
one encounters diagrams in which the produced particle itself couples to a
gauge boson $ B $, which leads to  
additional terms in the matrix elements.
In par\-tic\-u\-lar, the exchanged particle can be  an
$ e _ { \rm R } $ which can become soft in the
$ t $-channel. This contribution has
to be treated separately.

\subsection{Hard momentum transfer} 
\label{s:hard} 
We first consider the case that 
the exchanged particles have hard momenta. Then 
the equilibration rate can be determined
via the Boltzmann equation \cite{Weldon:1983jn,Laine:2013vpa}. 
We can write the time derivative of the $ L _ { e \rm R } $ density as
\begin{align}
  \dot n _ { L _ { e \rm R } }
   =
   \int _ \vec k \frac{ \partial }{ \partial t } \left[ f _ { \vec k } 
   - \bar  f _ { \vec k }   \right]
\label{dLdt}
\end{align}
where $ f _ { \vec k } $ and $ \bar f _ { \vec k } $ are the occupation numbers
of right-handed electrons and positrons. 
We replace the time derivatives on the right-hand side by the collision
term for $ 2 \to 2 $ particle scattering.
It contains the occupancies of the participating particles 
in the form 
\begin{align}
   f_1 f_2 [ 1 \pm f_3 ] [ 1 - f _ { \vec k } ] 
   - [ 1 \pm f_1 ] [ 1 \pm f_2 ] f_3 f _ { \vec k }
   ,
\label{gainloss}
\end{align}
corresponding to  gain and loss term.
The upper and lower signs are for bosons and fermions, respectively.

All Standard Model particles are in kinetic equilibrium due to their gauge
interactions. Therefore their occupancies are determined by the temperature
and by the chemical potentials of the slowly varying charges and of the
strictly conserved ones. 
To compute $ \GammaLL  $ at 
lowest order in chemical potentials, 
we can put all chemical potentials 
except $  \mu _ { e _ { \rm R } } $ equal to zero.
For the occupancy of right-handed electrons we can therefore write
\begin{align}
f _ { \vec k } 
   =    
   f _ { \rm F } ( k ^ 0    - \mu _ { e _ { \rm R } } )
\label{feR}
\end{align}
with 
$ k ^ 0 = | \vec k | $, 
$  \mu _ { e _ { \rm R } } = \mu _ { L _ { e \rm R } } $,
and for the other Standard Model
particles
\begin{align}
f _ i = f _ { \rm B, F } ( p _ i ^ 0 )
\label{fi}
.
\end{align}
In thermal equilibrium the gain and the loss term 
cancel,
\begin{align}
   f_1 f_2 [ 1 \pm f_3 ] [ 1 - f _ { { \rm F} } ( k ^ 0 )  ] 
   - [ 1 \pm f_1 ] [ 1 \pm f_2 ] f_3 f _ { { \rm F } } ( k ^ 0 ) 
   = 0
,
\label{boltz}
\end{align}
so that the collision term vanishes.
Expanding to first order in $ \mu _  { e _ {\rm R } } $ and making use of
(\ref{boltz}) together with
\begin{align}
f ' _ { \rm F } = - \frac 1 T
   f _ { \rm F } [ 1 - f _ { \rm F } ]
,
\label{fprime}
\end{align}
the contribution to the rate coefficient becomes 
\begin{align}
 \GammaLL ^ { 2 \to 2 , { \rm hard }  } 
    =
\frac  2 T
 \sum \limits _ { \rm processes }
 \int _ { {\vec k}, {\vec p} _ 1 ,
 {\vec p} _ 2 , {\vec p} _ 3 } 
  \frac { \sum   | \mathscr M | ^ 2  } 
  { 16 p _ 1 ^ 0 \, p _ 2 ^ 0 \, p _ 3 ^ 0 \, k ^ 0 } 
   \, ( 2 \pi ) ^ 4 & \delta ^ { ( 4 ) } ( p _ 1 + p _ 2 - p _ 3 - k ) 
 \nonumber
\\
 & 
 f _ 1 f _ 2 [ 1 \pm f _ 3 ]
 [ 1 - f _ { \rm F } (  k ^ 0 ) ] 
  \Big | _ { \rm hard } 
\label{G222}
   .
\end{align}
Both terms on the right-hand side 
of~\eqref{dLdt} give the same contribution which gives rise to
the factor $ 2 $.

One can write (\ref{G222}) in terms of the $ e _ { \rm R } $-production 
rate 
at vanishing $ e _ { \rm R } $-density,
\begin{align}
\GammaLL ^ { 2 \to 2 , { \rm hard } } 
=
\frac 2T
\int _ { \vec k }
\left [ 1 - f _ { \rm F } ( k ^ 0) \right ]
(2\pi)^3 
   \left .
   \frac { d  n _ { e _ { \rm R } } }
 {d t d ^ 3 k }
  \right | _ { n _ { e _ { \rm R } } =0 , \, { \rm hard } }
   ,
\label{gammafeq}
\end{align}
which is closely related to 
the production rate of sterile neutrinos computed in \cite{Besak:2012qm}. 
The difference between the two processes is that the sterile
neutrinos have no Standard Model gauge interactions, and therefore
do not interact once they are produced (at LO in their Yukawa couplings).
In contrast, the right-handed electrons carry weak hypercharge. 
Scatterings mediated by soft hypercharge gauge bosons contribute
to the LO rate, as discussed in section~\ref{s:lpm}. However for the
$ 2 \to 2 $ scattering of hard particles the soft scattering 
is a higher order effect and can be neglected here.

The diagrams contributing to the $ { e _ { \rm R } } $-production
are shown in figure~\ref{f:2to2diag}. The matrix elements for the 
processes with 
quarks and $ W $ bosons can be read off from \cite{Besak:2012qm} by
setting $ g' \to 0 $,
\begin{align}
\text{quarks}:
\label{Mquark}
    &     \qquad  \Sigma | \mathscr M | ^ 2
 = 6 \, h _ t  ^ 2 \, h _ e  ^ 2
 , \\
W \overline \varphi \rightarrow \overline \ell e _ { \rm R }:\label{WphileR}
    &     \qquad  \Sigma | \mathscr M | ^ 2
 = 3 g^2 \, h _ e  ^ 2 \, \frac u t 
  ,\\
W \ell \rightarrow \varphi e _ { \rm R }:\label{WlphieR}
    &     \qquad  \Sigma | \mathscr M | ^ 2 
 = 3 g^2 \, h _ e  ^ 2 \, \frac { - u } s 
   , \\
\ell \overline \varphi \rightarrow W e _ { \rm R }
	:\label{lphiWeR}
    &  
   \qquad  \Sigma | \mathscr M | ^ 2 	
  = 
   3 g^2 \, h _ e  ^ 2 \, \frac s { - t } 
  ,
\end{align} 
where~\eqref{Mquark} holds for any of the processes
$ \overline t Q _ 3 \rightarrow \ell e _ { \rm R }, \,
Q _ 3 \ell \rightarrow t e _ { \rm R }, \,
\overline t \ell \rightarrow \overline Q _ 3 e _ { \rm R }$.
For the processes with hypercharge gauge bosons we find
\begin{align} 
B \overline \varphi \rightarrow \overline \ell e _ { \rm R }:\label{BphileR}
    &     \qquad  \Sigma | \mathscr M | ^ 2
    = {g'}^2 \, h _ e  ^ 2 \, 
    \left[ 4 + \frac u t + \frac { 4t } u \right] 
   , \\
B \ell \rightarrow \varphi e _ { \rm R }:\label{BlphieR}
    &     \qquad  \Sigma | \mathscr M | ^ 2 
    = {g'}^2 \, h _ e  ^ 2 \,
        \left[ - 4 + \frac {-u} s + \frac { 4s } {-u} \right] 
   ,\\
\ell \overline \varphi \rightarrow B e _ { \rm R }:\label{lphiBeR}
    &     \qquad \Sigma | \mathscr M | ^ 2
    = {g'}^2 \, h _ e  ^ 2 \,
    \left[ - 4 + \frac s {-t} + \frac { 4 (-t) } s \right]
   .
\end{align}
Here we have summed over polarizations, color and weak isospin.
The Boltzmann equation can now be integrated as in 
\cite{Besak:2012qm}, with some
additional integrals due to the terms containing a factor $ 4 $ in
\eqref{BphileR} through~\eqref{lphiBeR}. We rewrite the contributions 
proportional
to $ 1 / ( - u ) $ as a process 
proportional to $ 1 / ( - t ) $ by interchanging the
incoming particles, which may change the statistics of particles 1 and 2.

In the integrals describing lepton exchange in $ t $-channel we
need to handle the infrared divergence appearing when 
the momentum of the exchanged lepton becomes small.
We proceed as  in \cite{Besak:2012qm} by introducing a transverse momentum cutoff
 $ q _ { \rm cut } $ 
for the exchanged particle with
$ gT \ll q _ { \rm cut } \ll T $. We isolate the piece which is singular
for $ q _ { \rm cut } \to 0 $ and integrate it analytically. Its 
logarithmic $ q _ { \rm cut } $ dependence drops out when combined with
the soft contribution (see section~\ref{s:soft}) which includes
only transverse momenta less than $ q _ { \rm cut } $. The remaining 
finite integral is then computed numerically. 

 \begin{figure}[t]
 \centering
 \hspace{.05\textwidth}
 \includegraphics[width=.35\textwidth]{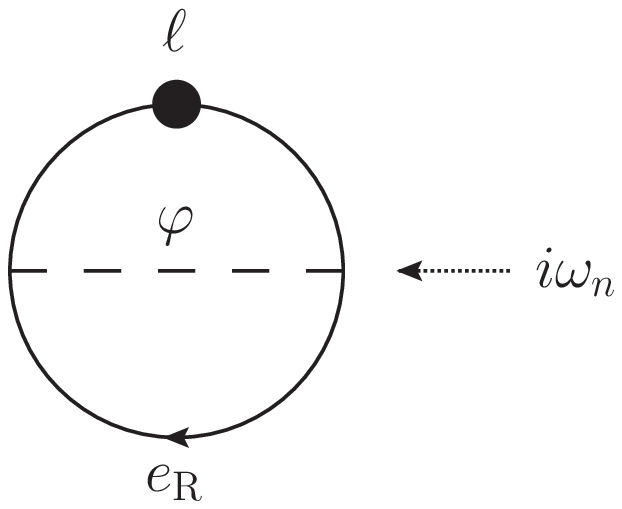}
 \hspace{.14\textwidth}
 \includegraphics[width=.35\textwidth]{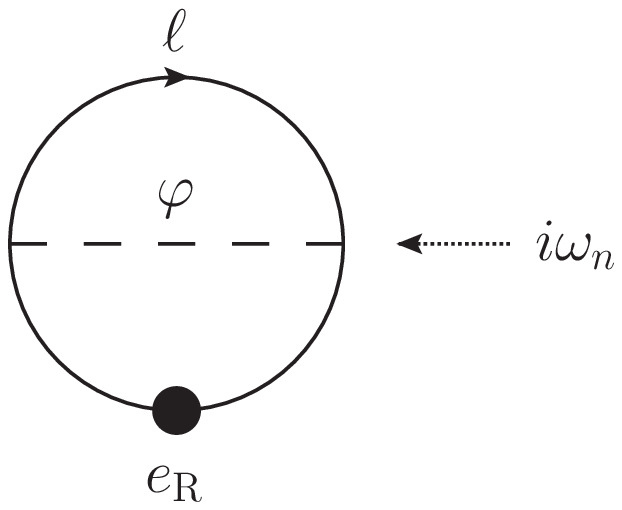}
 \hspace{.1\textwidth}
 \caption{Imaginary time correlator of the time derivative 
          of $ L _ { e \rm R } $ with one soft fermion. The 
          corresponding propagator has to be HTL resummed.
          The diagrams with $ {i \omega _ n \rightarrow - i \omega _ n } $
          are not shown.}
 \label{f:HTL}
 \end{figure}

\subsection{Soft momentum transfer}
\label{s:soft} 

The soft contribution is obtained from the retarded 
correlator using~\eqref{Gamma},
where either of the lepton propagators is HTL resummed. The corresponding 
diagrams are
shown in figure~\ref{f:HTL}. A straightforward computation in imaginary 
time, in which we make use of the sum rule found in \cite{Besak:2012qm},
and analytic continuation to real frequency
leads to 
\begin{align}
\GammaLL ^ { {\rm soft}  } 
=
\frac { h _ e  ^ 2 \, T }
      {  64 \, \pi } \,
   \left [ 
m _ \ell  ^ 2
\log \left( \frac { q _ { \rm cut } } { m _ \ell  } \right)
   +
m _ { e _ { \rm R } } ^ 2
\log \left( \frac { q _ { \rm cut } } { m _ { e _ { \rm R } } } \right)
   \right ] 
.
\label{softeR}
\end{align}

\subsection{Complete $ 2 \rightarrow 2 $ rate}
Adding
the hard singular and finite as well as the soft contributions, 
$ q _ { \rm cut } $ drops out, and the contribution from $ 2 \to 2 $ scatterings
to the rate coefficient $ \GammaLL$ is 
finite.
Evaluating the remaining integrals numerically, we find
\begin{align}
\GammaLL ^ { 2 \rightarrow 2 } 
=
\frac{ h _ e  ^ 2 \, T ^ 3 }
  { 2048 \,  \pi }
   \Bigg \{
      h _ t  ^ 2  c _ t 
      + \left ( 3 g ^ 2 + {g '} ^ 2 \right ) 
      &
         \left[ 
              c _ \ell 
              + \log \frac { 1 } { 3 g ^ 2 + {g '} ^ 2 }                       
         \right]
   \nonumber \\
      {}  + \, 
   4 {g '} ^ 2   
   &
   \left[ c _ { e _ { \rm R } }
                         + \log \frac { 1 } { 4 {g '} ^ 2 } \right]
\Bigg\}  
\label{gam2to2}
\end{align}
with
\begin{align}
c _ t                 = 2.82, \quad\quad\quad
c _ \ell              = 3.52, \quad\quad\quad
c _ { e _ { \rm R } } = 2.69.
\label{c}
\end{align}

\section{Results and discussion}
\label{s:res}

\begin{figure}[t]
\centering
 \includegraphics[width=.8\textwidth]{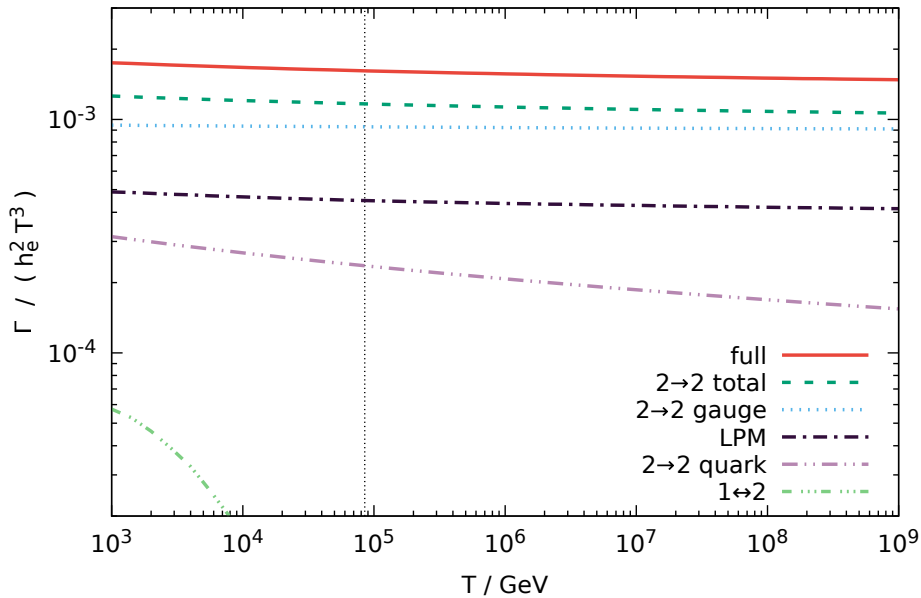}
 \caption{The rate coefficient $ \GammaLL $
   as function of the temperature.
    The curve labeled `full' incorporates all leading order contributions,
    `$2 \rightarrow 2 $~total' shows the full result  
    of~\eqref{gam2to2} whose contributions by gauge and
    quark scattering we show as `$2 \rightarrow 2 $ gauge' and `$2 \rightarrow 2 $ quark,'
    respectively. The curve labeled `LPM' shows the result
    of~\eqref{LPMrateP} and is the sum of the resummation of
    $1n \leftrightarrow 2n$ scatterings by soft gauge boson exchanges
    and the (inverse) Higgs decay labeled `$1\leftrightarrow 2$'.
    The dotted vertical line denotes the equilibration temperature~\eqref{Tstar}.
   }
 \label{f:gamma}
 \end{figure}
%
 \begin{table}[t]\centering
  \caption{Numerical values of the contributions from $ 2 \to 2 $ and LPM resummed multiple soft
           scattering to the equilibration rate coefficient $ \Gamma $. The numerical uncertainty
           in the LPM contribution is below 2~\%.}\vspace{3mm}
 \begin{tabular}{|c||c|c|c|}
  \hline
  $ T /\text{GeV}$ & $ \Gamma ^ { 2 \to 2 } / ( h _ e ^ 2 T ^ 3 ) $ 
                                       &  $ \Gamma ^ { \rm LPM } / ( h _ e ^ 2 T ^ 3 ) $ 
                                       &  $ \Gamma / ( h _ e ^ 2 T ^ 3 ) $ \\
  \hline
  \hline
$ 1.00 \cdot 10^{3} $ & $ 1.26 \cdot 10^{-3} $ & $ 4.89 \cdot 10^{-4} $ & $ 1.75 \cdot 10^{-3} $ \\
$ 4.00 \cdot 10^{3} $ & $ 1.22 \cdot 10^{-3} $ & $ 4.75 \cdot 10^{-4} $ & $ 1.70 \cdot 10^{-3} $ \\
$ 1.60 \cdot 10^{4} $ & $ 1.19 \cdot 10^{-3} $ & $ 4.62 \cdot 10^{-4} $ & $ 1.65 \cdot 10^{-3} $ \\
$ 6.40 \cdot 10^{4} $ & $ 1.17 \cdot 10^{-3} $ & $ 4.51 \cdot 10^{-4} $ & $ 1.62 \cdot 10^{-3} $ \\
$ 2.56 \cdot 10^{5} $ & $ 1.15 \cdot 10^{-3} $ & $ 4.43 \cdot 10^{-4} $ & $ 1.59 \cdot 10^{-3} $ \\
$ 1.02 \cdot 10^{6} $ & $ 1.13 \cdot 10^{-3} $ & $ 4.36 \cdot 10^{-4} $ & $ 1.57 \cdot 10^{-3} $ \\
$ 4.10 \cdot 10^{6} $ & $ 1.11 \cdot 10^{-3} $ & $ 4.31 \cdot 10^{-4} $ & $ 1.54 \cdot 10^{-3} $ \\
$ 1.64 \cdot 10^{7} $ & $ 1.10 \cdot 10^{-3} $ & $ 4.26 \cdot 10^{-4} $ & $ 1.53 \cdot 10^{-3} $ \\
$ 6.55 \cdot 10^{7} $ & $ 1.09 \cdot 10^{-3} $ & $ 4.22 \cdot 10^{-4} $ & $ 1.51 \cdot 10^{-3} $ \\
$ 2.62 \cdot 10^{8} $ & $ 1.07 \cdot 10^{-3} $ & $ 4.18 \cdot 10^{-4} $ & $ 1.49 \cdot 10^{-3} $ \\
$ 1.05 \cdot 10^{9} $ & $ 1.06 \cdot 10^{-3} $ & $ 4.14 \cdot 10^{-4} $ & $ 1.47 \cdot 10^{-3} $ \\
  \hline
 \end{tabular}
 \label{t:num}
 \end{table}
For our numerical results we evaluate the 1-loop running couplings at 
the renormalization scale $ \pi  T $. We have checked that 
increasing the renormalization scale by a factor 2 changes
our results by less than 3\% in the entire temperature range 
we consider.

Figure~\ref{f:gamma} shows
the various contributions to the equilibration rate. 
The $ 2 \rightarrow 2 $ processes are dominant over the entire
temperature range considered. The largest contribution comes from
scatterings off hard gauge bosons. 
The $ 1n \leftrightarrow 2n$ contribution
is about a factor $ 0.4 $
smaller than the total $ 2 \to 2 $ rate.
Except at very low
temperature the (inverse) Higgs decay gives a negligible contribution,
and it vanishes completely above $ T \simeq 60 $ TeV. 
In table~\ref{t:num} we show numerical values for the total $ 2 \to 2 $
as well as the LPM resummed contribution along with the full result for $ \Gamma $.

The LPM resummed rate is a complicated function of the coupling
constants and there is not such a simple expression like~(\ref{gam2to2}) 
for the $ 2 \to 2 $ rate. Inspired by the form of~(\ref{gam2to2}) we have
fitted the LPM contribution with a similar expression,
\begin{align}
\GammaLL ^ { \text{LPM} } 
\approx
\frac{ h _ e  ^ 2 \, T ^ 3 }
  { 2048 \,  \pi }
\left \{
h _ t  ^ 2  d _ t 
+ ( 3 g ^ 2 + {g '} ^ 2 ) d _ \ell
+ 4 {g '} ^ 2  d _ { e _ { \rm R } }
\right \}
\label{LPMfit}
   .
\end{align}
We find that with 
\begin{align}
d _ t                 = 1.48, \quad\quad\quad
d _ \ell              = 0.776, \quad\quad\quad
d _ { e _ { \rm R } } = 2.03
\label{d}
\end{align}
the relative error of $ \GammaLL ^ { \rm LPM } $
is much smaller than our numerical uncertainty
throughout the temperature range 
$ 10 ^ 3 \text{ GeV} \leq T \leq 10 ^ 9 \text{ GeV} $.

The right-handed electron lepton number
comes into equilibrium around the temperature
$ T _ { \rm eq } $ 
at which $ \gammaLL  $ equals  
the Hubble rate
\footnote{In \cite{Cline:1993bd} a different definition of the
$ e _ { \rm R } $ equilibration temperature is used.}
\begin{align}
   H 
   = 
   \sqrt{ \frac { 4 \pi ^ 3 g _ * } { 45 } } \frac { T ^ 2 } { m _ { \rm Pl } }
\label{hubble}
.
\end{align}
Here $ g _ *  $ is the number of relativistic degrees of freedom 
with   $ g _ * = 106.75 $ in the Standard Model,
$ m _ { \rm Pl } = 1.22 \cdot  10 ^{ 19 } $ GeV is
the Planck mass. Using~\eqref{gGexplicit}
we find for the equilibration 
temperature $T_ { \rm eq}$ of the right-handed electron lepton
number in the Standard Model
\begin{align}
T_{ \rm eq }  = 8.5 \cdot 10^4 \text{ GeV}
   .
\label{Tstar}
\end{align}
This value lies in the temperature region  in which 
leptogenesis through neutrino oscillations \cite{Akhmedov:1998qx,Asaka:2005pn} 
can take place, see \emph{e.g.}
\cite{Hernandez:2016kel,Ghiglieri:2017csp}.
In this case 
$ X _ \alpha $ and 
$ L _ { e \rm R } $ are violated on 
similar time scales, and the kinetic equations must describe
the evolution of all four quantities.

It is interesting to see how hypercharge gauge interactions affect the
$ e _ { \rm R } $-equilibration, since they give rise to diagrams which are not
present in sterile neutrino production. We find that they substantially 
boost the
equilibration rate. In table~\ref{t:gprime} we
show the increase in the complete rate compared to the
result with  $ g ' =0 $.  Despite the relative
smallness of $ g' $, its effect on the equilibration rate is quite
significant, and it increases with the temperature
due to the different running of $ g ' $ and $ g $.
 
The first calculation of the $ e _ { \rm R } $-equilibration
rate was performed in~\cite{Campbell:1992jd}, where only 
the $ 2 \to 1 $ inverse Higgs decay
is taken into account
and 
thermal fermion masses as well as
the final state distribution function are 
neglected.
At ${T = 10^3\text{ GeV}}$ our result is about 5 times as large
as the one obtained in~\cite{Campbell:1992jd}.
Around the equilibration temperature~\eqref{Tstar}
the inverse decay is not even kinematically allowed 
when thermal fermion masses are included, 
here we obtain a result that is about 6 times the one obtained
by the approximations of~\cite{Campbell:1992jd}.

 \begin{table}[t]\centering
  \caption{Relative increase of   $ \GammaLL $ 
           when hypercharge gauge interactions are included.}\vspace{3mm}
 \begin{tabular}{|c||c|c|c|}
  \hline
  $ T $          & $10^3\text{ GeV}$ & $ 10^6\text{ GeV} $  & $10^9\text{ GeV}$   \\
  \hline
  \hline
  LPM            &     21\%          &       30\%           &      39\%           \\
  $2\to 2 $      &     34\%          &       44\%           &      53\%           \\
  \hline
  total          &     30\%          &       40\%           &      49\%           \\
  \hline
 \end{tabular}
 \label{t:gprime}
 \end{table}

Reference \cite{Cline:1993bd} includes
$ 2 \leftrightarrow   2 $ processes  as well as the (inverse) Higgs decays 
while neglecting $ 1n \leftrightarrow  2n $ scattering. 
We can compare the $ 2 \to 2 $ scattering
rates involving quarks. Therefore we recompute $ c _ t $ in~\eqref{gam2to2}
using Maxwell-Boltzmann statistics for all particles, leading to
$ c _ t ^ { \rm MB }  = 2.14 $, which is a relative error of $ 24\% $ 
compared to the correct
quantum statistics, as anticipated in \cite{Cline:1993bd}. Our result for classical
statistics is
$ 9\% $ larger than the one obtained in \cite{Cline:1993bd}.
We can also compare the gauge contribution to the $ 2 \to 2 $ scatterings. 
With the values for the gauge couplings
of \cite{Cline:1993bd}, our result is about $ 50\% $ larger 
\footnote{Cf.~equations (25) through (27) in \cite{Cline:1993bd}.}
which could be due to the use of classical statistics and of
zero-momentum thermal fermion masses in \cite{Cline:1993bd}.

The equilibration 
of right-handed muons and taus 
in a temperature regime between $10^7$ 
and $ 10^{13}\text{ GeV} $ is considered in
\cite{Garbrecht:2013urw}, by including the (inverse) Higgs decays
and $ 2 \leftrightarrow  2 $ scatterings.
By removing the inverse susceptibilities and the slow Yukawa couplings,
we can compare our results for
$ \GammaLL / h _ e ^ 2 $, because it is lepton-flavor independent.
We find our full rate to be 
$ 2.8 $ times their result.  The authors also estimate the
effect of multiple soft scattering.%
\footnote{See equation (97) in \cite{Garbrecht:2013urw}.}  The 
relative magnitude of the effect
of multiple soft scattering is estimated in
\cite{Garbrecht:2013urw} as 
${\gammaLL ^ { \rm LPM } / \gammaLL ^ { 2 \to 2 } \sim
  0.25}$, while we obtain about~$ 0.4 $. Our result
for the quark contribution to $ 2 \to 2 $ scattering is $ 2 $~times
the result in \cite{Garbrecht:2013urw}, and both our logarithmic
contributions to $ \GammaLL ^ { 2 \rightarrow 2 }  $ are $ 2.2 $~times as large as the ones
in~\cite{Garbrecht:2013urw}.

To summarize, we have computed the equilibration
rate of right-handed electrons in the symmetric phase
by including, for the first time, all Standard
Model processes at leading order in the couplings. We have found that
the dominant processes are $ 2 \to 2 $ scatterings. Leading order contributions
are also given by inverse Higgs decays and additional soft scattering which
was included by Landau-Pomeranchuk-Migdal~(LPM) resummation.
We obtain an equilibration rate which is substantially larger than 
approximations  presented in previous literature.
Our result shows that it can be important to include the process of 
$ e _ { \rm R } $ equilibration in low-scale leptogenesis.

\bigskip 

\noindent {\bf  \large Acknowledgments } We would like to thank Peter~Arnold for
sharing his insights into the LPM effect, Mikko Laine for valuable comments
on the manuscript, Guy Moore and S\"oren Schlichting
for useful discussions, and Andrew Long and Eray Sabancilar for correspondence.  
This work was funded in part by
the Deutsche Forschungsgemeinschaft (DFG, German Research
    Foundation) – Project number 315477589 – TRR 211.

\appendix
\global\long\def\theequation{\thesection.\arabic{equation}}

\section{Susceptibilities}
\label{a:sus}
In this appendix we compute the susceptibilities which relate
the chemical potentials to the densities of slowly violated and of 
strictly conserved charges in~\eqref{na} at leading (zeroth) order in the
couplings.
When the right-handed electrons come into
equilibrium, the $ \mu $ and $ \tau $ Yukawa interactions are already fast, 
so that the lepton numbers carried by the corresponding
right-handed particles are not conserved. Having expanded already in $ h _ e $,
the violation of $ L _ { e \rm R }  $ is of higher order and we can 
consider it conserved,
hence we introduce a chemical potential $ \mu _ { L _ { e \rm R } } $.
The weak hypercharge $ Y $ is strictly conserved.
The zero-momentum 
mode of the temporal component of the hypercharge gauge field
$ B _ 0 $ plays the role of the corresponding
chemical potential, see (\ref{muY}).
Integrating over $ B _ 0 $ enforces $ Y =0 $.
Then the particle chemical potentials read
\begin{align}
\mu _ { Q } 
  &= 
   \frac 16 \mu _ Y - \frac 13 \mu _ X , & 
              \mu _ { u _ { \rm R } } 
   &= 
   \frac 23 \mu _ Y - \frac 13 \mu _ X , &
              \mu _ { d _ { \rm R }  } 
   &= - \frac 13 \mu _ Y - \frac 13 \mu  _ X              \nonumber \\
\mu _ { \ell _ \alpha } 
   &= - \frac 12 \mu _ Y + \mu _ { X _ \alpha }, &
              \mu _ { e _ { \rm R } } 
   &= -\mu _ Y + \mu _ { X _ e } + \mu _ {L _ { e \rm R } }, &
              \mu _ { \mu _ { \rm R } } 
   &= -\mu _ Y + \mu _ { X _ \mu }   \nonumber \\
\mu _ { \tau _ { \rm R } } 
   &= -\mu _ Y + \mu _ { X _ \tau }, &
              \mu _ { \varphi } 
   &= \frac 12 \mu _ Y
   , 
   &  & \label{mupart}
\end{align}
with $ \mu _ X \equiv \frac 13 \sum _ { \alpha } \mu _ { X _ \alpha } $.
Now we compute the pressure and obtain the
matrix of susceptibilities via~\eqref{chiPmu}. Inversion of this matrix yields
\begin{align}
\chi ^ { - 1 } =
\frac { 1 } { 481 \, T ^ 2 }
\left(
\begin{array}{ccccc}
 4266 & -1110 & 312 & 312 & 270 \\
 -1110 & 1332 & 0 & 0 & 222 \\
 312 & 0 & 1066 & 104 & 312 \\
 312 & 0 & 104 & 1066 & 312 \\
 270 & 222 & 312 & 312 & 492 \\
\end{array}
\right)
\label{chiinveR}
\end{align}
for 
the ordering $\left\{ L _ { e \rm R }, X _ e, X _ \mu, X _ \tau, Y \right\} $.

\section{Solving the integral equation}
\label{a:sol}

The Fourier transformation 
\begin{align}
\vec f ( \vec B ) \equiv \int \frac{ d ^ 2 P }{ ( 2 \pi ) ^ 2 } 
      e ^ { i \vec P \cdot \vec B } \, { \vec f } ( \vec P )
\label{fourier}
\end{align}
turns the integral equation (\ref{intf}) 
for $ \vec f ( \vec P ) $ into a differential equation 
for $ \vec f ( \vec B ) $,
\begin{align}
i \beta \left( \Delta - M ^ 2 \right) \vec f ( \vec B )
=
\mathscr K ( B )\,  \vec f ( \vec B ) - 2\, i \, \nabla \delta ^ { ( 2 ) } ( \vec B ),
\label{fpde}
\end{align}
where the differential operators act on the two-dimensional impact
parameter $ \vec B $. We denote $ B \equiv | \vec B | $ and we have introduced
\begin{align}
\mathscr K ( B )
&\equiv
\frac { 3 g ^ 2 T } { 4 } \mathscr D ( x _ k m _ {\rm D} B )
\nonumber 
   \\
&+
 { g ' } ^ 2 T  
   \big[
   y _ \varphi y _ \ell \,  \mathscr D( x _ k m _ {\rm D} ' B ) 
   + y _ \ell y _ { e _ { \rm R } } \, \mathscr D \left( m _ {\rm D} ' B \right)
             - y _ \varphi  y _ { e _ { \rm R } } 
         \, \mathscr D \left( x _ p m _ {\rm D} ' B \right )
    \big]
   \label{K}
\end{align}
with
\begin{align}
\mathscr D ( y )
\equiv \frac { 1 } { 2 \pi } \left[ 
\gamma _ { \rm E } + K _ 0 ( | y | ) + \log \left| \frac y2 \right|
\right] 
   .
\label{D}
\end{align}
$ \gamma _ { \rm E } $ is the Euler-Mascheroni constant and $ K _ 0 $ is a
modified Bessel function.
In terms of the Fourier transform the real part in
\eqref{LPMrateP} becomes
\begin{align}
{ \rm Re } 
\int \frac { d ^ 2 P } { ( 2 \pi ) ^ 2 }
\vec P \cdot \vec f ( \vec P )
=
\lim \limits _ { \vec B \rightarrow \vec 0 } { \rm Im } \nabla \cdot { \vec f } ( \vec B )
\label{intlimrel}
   .
\end{align}
Writing $ \vec f ( \vec B ) \equiv \vec B \, h (  B  ) $, 
we arrive at the following ordinary differential equation
for $ h ( B ) $, valid at $ B \neq 0 $,
\begin{align}
i \beta 
\left\{
\frac { d^ 2 } { d B ^ 2 } + \frac 3B \frac { d } { d B } - M ^ 2
\right\} 
h ( B ) 
-
\, \mathscr K ( B ) \, 
h ( B ) = 0
   .
\label{ODE}
\end{align}
In terms of $ h $, the relation~\eqref{intlimrel} becomes
\begin{align}
{ \rm Re } 
\int \frac { d ^ 2 P } { ( 2 \pi ) ^ 2 }
\vec P \cdot \vec f ( \vec P )
=
2 \lim \limits _ { B \rightarrow 0 } { \rm Im } \, h ( B ).
   \label{Reh}
\end{align}
For 
$ B \rightarrow 0 $ the function  $ h $ 
has a singularity which is determined by 
the delta function in~\eqref{fpde},
\begin{align}
h ( B ) \overset{ B \rightarrow 0 } { \sim } -\frac { 1 } { \pi \beta B^2 }
\label{hlimit}
   ,
\end{align}
and which is insensitive to $ \mathscr K  $.
Being purely real, this singularity does not enter~\eqref{Reh}. 
We write $ h = h ^ { \rm decay }+  h ^{ \rm scat } $,
where $ h ^ { \rm decay } $ 
contains only the (inverse) Higgs decay contribution. We obtain
it by solving 
\eqref{intf} with
$ \int   d ^ 2 q _ \perp \{ \cdots \}
\rightarrow 0 ^ + \, 
\vec f ( \vec P )  $
and then taking the Fourier transform. This gives
\begin{align}
h ^ { \rm decay}  ( B ) 
   = 
   \left \{ 
\begin{array}{lll}
-\dfrac { m } { \pi \beta B } K _ 1 ( m B )
   & 
   ( M ^ 2 > 0 ) 
   \\ [\bigskipamount]
 \dfrac { m } { 2 \beta B } 
   \left[ Y _ 1 ( m B ) 
   - i \, { \rm sign } ( \beta ) \, J _ 1 ( m B )
\right]
   &  
   ( M ^ 2 < 0 ) 
\end{array}
   \right .
   \label{hdecay}
\end{align}
with $   m \equiv \sqrt{ | M ^ 2 | } $, 
and the (modified) Bessel functions $ K _ 1, Y _ 1 $ and $ J _ 1 $.
Then we solve the  differential equation for $ h ^ { \rm scat } $ 
numerically as described in \cite{Anisimov:2010gy}.

\section{Integrals appearing in the $ 2 \rightarrow 2 $ rate}
\label{a:I}

After summing over all leading order processes the production rate
on the right-hand side of
\eqref{gammafeq} can be written as 
\begin{align} 
\left .
   \frac { d  n _ { e _ { \rm R } } }
 {d t d k ^ 0 }
  \right | _ { n _ { e _ { \rm R } } =0 }
 =
  \frac { h _ e  ^2 f _ { \rm F } ( k ^ 0 ) } { 128 \pi ^ 5 } 
  \Big[
& 18 \, h _ t  ^ 2 \, \mathscr I _ { \rm f f f } ^ 0 
   \nonumber \\
+ & ( 3 g ^ 2 + {g '} ^ 2 ) \left\{ \mathscr I _ { \rm b f b } ^ 1
                                  + \mathscr I _ { \rm b b f } ^ 1
                                  + \mathscr I _ { \rm f b b } ^ 1 \right\}
                                   \nonumber \\
+ & 4 { g ' } ^ 2 \left\{ \mathscr I _ { \rm b f b } ^ 1
                        + \mathscr I _ { \rm b b f } ^ 1
                        + \mathscr I _ { \rm f b b } ^ 1
                        + \mathscr I _ { \rm b b f } ^ 0
                        - 2 \mathscr I _ { \rm b f b } ^ 0  \right\} \Big]
\label{gammaI}
\end{align}
with $ k ^ 0 = | \vec k | $. Here 
we have already integrated over the direction of $\vec k$.
The  $ \mathscr I _ {
  1 2 3 } ^ n $ are the different 
phase space integrals appearing in (\ref{G222}). 
The lower indices refer to the
statistics of the particles $ 1, 2, 3 $ and the upper index $ n =0,1$
is 
the power of  the ratios of Mandelstam
variables in equations (\ref{Mquark})-(\ref{lphiBeR}).
The exact definitions of the $ \mathscr I $ are given below.

Like in \cite{Besak:2012qm} 
we carry out some integrations 
analytically until there are two integrals over the variables
$ q _ \pm \equiv ( q ^ 0 \pm
|\vec q| ) / 2 $ left. If not stated otherwise,
$ q $ is the exchanged 4-momentum.  
For each process we decompose the products of occupancies 
in~\eqref{G222} as
\begin{align}
 f _ 1 f _ 2 [ 1 \pm f _ 3 ]
 =
 f _ { \rm F } ( k ^ 0) \widetilde f \widehat f
 \label{decomp}
 ,
\end{align}
where $ \widehat f $ is a function of $ q _ + + q _ -$
and of the energy of one incoming  particle only. 
Most of the integrals appear in sterile neutrino production as
well. For the sake of completeness, we list  them in this
appendix, adopted to our notation. We also give the
analytic integrals which were not computed in \cite{Besak:2012qm}.
The terms containing $ 1 / t $ are 
infrared divergent when integrated over $ q _ \pm $. 
All divergent contributions encountered here already appear in sterile  
neutrino production (see \cite{Besak:2012qm} for details). 

\subsection{$ \mathscr I _ { \rm f f f } $}
This integral is exclusive to quark scattering. Since the
squared matrix elements do not depend on the Mandelstam
variables, we may
choose $
q = p _ 3 + k $ for both  $ s $- and 
$ t $-channel.  We find
\begin{align}
\widetilde f &= f _ { \rm B } ( q _ + + q _ - ) + f _ { \rm F } ( q _ + + q _ - - k ^0 )\\
\widehat f &= 1 - f _ { \rm F } ( q _ + + q _ - - E _ 2 ) - f _ { \rm F } ( E _ 2 )
,
\end{align}
and we have
\begin{align}
\mathscr I _ { \rm f f f } ^ 0 = 
  \int \limits _ { k ^ 0 } ^ \infty \! d q _ + \int \limits _ 0 ^ { k ^ 0 } \! d q _ -
  \, \widetilde f
    \int \limits _ { q _ - } ^ { q _ + } \! d E _ 2 \, \widehat f 
\label{Ifff}
.
\end{align}
Only $ n = 0 $ appears, and the integral of $ \widehat f $ over $ E _
2 $ is given by equation (A.10) of \cite{Besak:2012qm}.

\subsection{$ \mathscr I _ { \rm b f b } $}

This integral appears in $ s $-channel processes, so that 
$ q = p _ 3 + k $.  We have
\begin{align}
  \widetilde f &= f _ { \rm F } ( q _ + + q _ - )
  + f _ { \rm B } ( q _ + + q _ - - k ^ 0 )
   \label{ftbfb}
   \\
\widehat f 
   &= 1 + f _ { \rm B } ( q _ +  + q _ - - E _ 2 ) 
   - f _ { \rm F } ( E _ 2 )
   \label{fhbfb}
,
\end{align}
and we need 
\begin{align}
\mathscr I _ { \rm b f b } ^ n = 
\int \limits _ { k ^ 0 } ^ \infty \! d q _ + \int \limits _ 0 ^ { k ^ 0 } \! d q _
  - \, \widetilde f
    \int \limits _ { q _ - } ^ { q _ + } \! d E _ 2 \, \widehat f 
    \left( \frac { \langle -u \rangle } { s } \right) ^ n
\label{Ibfb}
\end{align}
where $ { \langle -u \rangle } $ is the Mandelstam variable $ u $ averaged over angles,
\begin{align}
\frac { \langle -u \rangle } { s } = 
\frac { q _ + ^ 2 + q _ - ^ 2 - ( q _ + + q _ -) ( E _ 2 + k ^ 0 ) + 2 E _ 2 k ^ 0 } 
{ ( q _ + - q _ - ) ^ 2 }
.
\label{uovers}
\end{align}
The result of 
the $ E _ 2 $ integration is found in equation (A.13) of \cite{Besak:2012qm}.
For the $ n = 0 $ integral we obtain
\begin{align}
\int \limits _ { q _ - } ^ { q _ + } \! d E _ 2 \, \widehat f 
=
- (q _ + - q _ - ) + T \left[ \log \left( - 1 + e ^ { 2 q _ + / T } \right)
                            - \log \left( - 1 + e ^ { 2 q _ - / T } \right) \right]
.                            
\end{align}
\subsection{$ \mathscr I _ { \rm b b f } $}
This function arises in $ t $-channel processes, so that $ q = p _ 1 - p _ 3 $.
We obtain 
\begin{align}
  \widetilde f &= 1 + f _ { \rm B } ( k ^ 0 - q _ + - q _ - )
  - f _ { \rm F } ( q _ + + q _ - )
  \label{ftilbbf}
  \\
\widehat f 
   &= 
   f _ { \rm B } ( E _ 1 ) + f _ { \rm F } ( E _ 1 - q _ + - q _ - )
   \label{fhbbf}
\end{align}
such that
\begin{align}
\mathscr I _ { \rm b b f } ^ n = 
    \int \limits _ 0 ^ { k ^ 0 } \! d q _ + \int \limits _ {-\infty} ^ 0 
    \! d q _ - \, \widetilde f
    \int \limits _ { q _ + } ^ { \infty } \! d E _ 1 \, \widehat f 
    \left( \frac { \langle u \rangle } { t } \right) ^ n
.
\label{Ibbf}
\end{align}
Here we have
\begin{align}
\frac { \langle u \rangle } { t } = 
\frac { 2 q _ + q _ - + 2 E _ 1 k ^ 0 - ( q _ + + q _ - )( E_1 + k ^ 0 ) } 
{ ( q _ + - q _ - ) ^ 2 }
.
\label{uovert}
\end{align}
The $ E _ 1 $ 
integral with $ n = 1 $ is equation (A.24) of \cite{Besak:2012qm}, 
while for the case
$ n = 0 $ we get
\begin{align}
\int \limits _ { q _ + } ^ { \infty } \! d E _ 1 \, \widehat f 
=
q _ + + q _ - 
+ T \left[ \log\left(1 + e^{- q _ - / T } \right) 
   - \log \left(-1 + e^{q_+/T}\right)
                  \right]
.                            
   \label{bbf}
\end{align}
\subsection{$ \mathscr I _ { \rm f b b } $}
We encounter this integral in $ t $-channel, so again $ q = p _ 1 - p _ 3 $.
Here
\begin{align}
  \widetilde f
  &= 
   1 + f _ { \rm B } ( k ^ 0 - q _ + - q _ - ) 
   - f _ { \rm F } ( q _ + + q _ - )
   \label{ftfbb}
  \\
\widehat f &= f _ { \rm F } ( E _ 1 ) + f _ { \rm B } ( E _ 1 - q _ + - q _ - )
   \label{fhfbb}
\end{align}
and we have
\begin{align}
\mathscr I _ { \rm f b b } ^ 1 = 
    \int \limits _ 0 ^ { k ^ 0 } \! d q _ + \int \limits _ {-\infty} ^ 0
    \! d q _ - \, \widetilde f
    \int \limits _ { q _ + } ^ { \infty } \! d E _ 1 \, \widehat f 
    \left( \frac { \langle s \rangle } { - t } \right)
.
\label{Ifbb}
\end{align}
We write $\langle s \rangle / ( - t ) = 1 + \langle u \rangle / t
$ with $ \langle u \rangle / t $ from~\eqref{uovert}.  We only need $
n = 1 $, and the corresponding integral over $ E _ 1 $ is found
in \cite{Besak:2012qm} in (A.20).

\section{Conversion of $ \LeR $ to hypercharge gauge fields}
\label{a:ins}
Even without Yukawa interaction the conservation of 
$ \LeR $ is violated by the chiral anomaly
\footnote{One can find different prefactors on the right-hand side of~\eqref{anomaly}
          in the literature, which are related to different conventions for the 
          weak hypercharge. Ours is the same as in~\cite{Kamada:2016eeb}.}
\begin{align}
\partial _ { \mu } j _ { e \rm R } ^ { \mu } 
=
-
\frac { y _ { e _ { \rm R } } ^ 2 { g ' } ^ 2 } {32 \pi ^ 2 }
\varepsilon ^ { \mu \nu \rho \sigma } F _ { \mu \nu } F _{ \rho \sigma }
\label{anomaly}
,
\end{align}
with
\begin{align}
j _ { e \rm R } ^ \mu 
\equiv
\overline { e _ { \rm R }} \, \gamma ^ \mu e _ { \rm R }
\label{jeR}
.
\end{align}
$ F _ { \mu  \nu  } $ denotes the hypercharge field strength, and 
we use the convention $ \epsilon  ^{ 0123 } = +1 $. This may lead
to interesting effects, such as the generation of primordial magnetic
fields \cite{Joyce:1997uy,Long:2013tha}. In this appendix
we want to see when the anomaly can affect  
the long time and large distance behavior of $ j _ { e \rm R } $.
Even if there are no gauge fields present initially, there is an instability
in the gauge fields for non-zero $ \mu  _ { \LeR } $, leading to exponential
growth \cite{Joyce:1997uy}. 
We compute the maximal growth rate of the unstable 
modes in order to derive a bound
on $ \left| \mu _  { L _  { e \rm R } } \right| $, below which the growth 
is smaller than the equilibration rate $ \gamma $ and can be neglected
in the kinetic equation~\eqref{kineqn}.

The hypercharge electric and magnetic fields $ \vec E $ 
and $ \vec B $ 
with wavelengths greater than the
particle mean free path
are described by 
magneto-hydrodynamics. 
In the presence of the anomaly (\ref{anomaly}), 
in addition to the usual ohmic current
$ \vec j _ { \rm Ohm } = \sigma  \vec E $ with the
hyperelectric conductivity $ \sigma  $, one has to take into account
the  contribution 
\cite{Son:2009tf,Joyce:1997uy} 
\begin{align}
   \vec j _ { \rm anomaly } = - \frac { y _ { e \rm R } ^ 2 { g' } ^ 2 } 
   { 4 \pi  ^ 2 }
   \mu  _ { \LeR } \vec B 
   \label{jano}
   .
\end{align} 
The fields evolve on time scales much larger than  $ \sigma  ^{ -1 } $.
Therefore $ \dot { \vec E } $ is much smaller than $ \sigma  \vec E $,
and can be neglected in the equations of motion which become
\begin{align}
\vec E 
   &= \frac 1\sigma\left[\nabla \times \vec B 
+ \frac { y _ { e \rm R } ^ 2 { g' } ^ 2 } 
   { 4 \pi  ^ 2 }
                \mu _ { L _ { e \rm R } }    \vec B \right]
                \label{MaxwellE}
, \\
   \dot { \vec B  } 
   &= - 
   \nabla \times \vec E
\label{MaxwellB}
   .
\end{align}
Using $ \nabla \cdot \vec B = 0$, these 
can be recast as
\begin{align}
   \dot { \vec B }   + \frac 1\sigma \left[ - \laplace \vec B 
                     + \frac { y _ { e \rm R } ^ 2 { g' } ^ 2 } 
   { 4 \pi  ^ 2 }
\mu _ { L _ { e \rm R } } 
                          \nabla \times \vec B \right]
            = \vec 0
            .
    \label{Bdot}
\end{align}
Following~\cite{Joyce:1997uy}, we Fourier transform $ \vec B (t,\vec x ) = \int _ { \vec k } \vec B _ { \vec k } (t) e ^ { i \vec k \cdot \vec x }$ to obtain
\begin{align}
\sigma \dot{  \vec B}  _ { \vec k } + \vec k^2 \vec B _ { \vec k }
               + i \frac { y _ { e \rm R } ^ 2 { g' } ^ 2 } 
   { 4 \pi  ^ 2 }
   \mu_{L _ { e \rm R } }
              \vec k \times \vec B _ { \vec k } = \vec 0
              \label{fourierB}
              .
\end{align}
Now decompose $ \vec B _ { \vec k } =\sum _{ i=1}^{2} b _ i \vec e _ i $. The $ \vec e _ i $
are an orthonormal basis in the plane orthogonal to $ \vec k $. 
The equations for $ b _ \pm \equiv b _ 1 \pm i b _ 2 $ decouple,
\begin{align}
   \sigma \dot b   _ \pm
   = -
   | \vec k | \left ( 
   | \vec k | \mp   
                   \frac { y _ { e \rm R } ^ 2 { g' } ^ 2 } 
   { 4 \pi  ^ 2 }
\mu _ { L _ { e \rm R } } 
   \right ) b _ \pm 
\label{circ}
.
\end{align}
For 
\begin{align}
| \vec k | < 
  \frac { y _ { e \rm R } ^ 2 { g' } ^ 2 } 
   { 4 \pi  ^ 2 }
   | \mu _ { L _ { e \rm R } } | 
\label{kinst}
\end{align}
there is an instability 
with the growth rate 
\begin{align}
\gamma _ { \rm inst} 
   = 
   \frac { | \vec k | } \sigma
   \left(   \frac { y _ { e \rm R } ^ 2 { g' } ^ 2 } 
   { 4 \pi  ^ 2 }
   | \mu _ { L _ { e \rm R } } | 
   - | \vec k | \right)
   .
   \label{ginst}
\end{align}
The maximal growth rate is
\begin{align}
   \gamma _ { \rm inst} ^ { \rm max}   
    =  
      \frac { \mu _ { \LeR } ^ 2 y _ { e \rm R } ^ 4 { g' } ^ 4}
     { 64 \pi  ^ 4 \sigma  }  
    \label{gmax}
         .
\end{align}
The linear kinetic equation neglecting the dynamics of 
the long wavelength hypermagnetic
fields is valid as long as the magnetic dynamics happen on longer time scales
than the perturbative ones characterized by the equilibration rate $ \gamma $,
\begin{align}
\gamma _ { \rm inst} ^ { \rm max} < \gamma
   \label{glim}
   .
\end{align}
In the 
leading logarithmic approximation the hyperelectric 
conductivity is  \cite{Arnold:2000dr}
\begin{align}
   \sigma 
   =
   C \frac { T } { {g'} ^ 2 \log ( 1 / g' ) }
,
   \label{conduc}
\end{align} 
with $ C = 7.05 $ in the Standard Model with one Higgs doublet.%
\footnote{ Adding further Higgs doublets decreases the conductivity.}  
At the equilibration temperature~\eqref{Tstar} in the 
Standard Model,~\eqref{glim}
translates into the condition (\ref{constraint}).

Consider again 
example 3
of section~\ref{s:rhe}. Here the constraint (\ref{constraint}) 
implies  
\begin{align}
\left|
Y _ { L _ { e \rm R } }
-
\frac {1110}{4266} Y _ { X _ e }
+\frac{312} {4266} \left( Y _ { X _ {\mu} } + Y _ { X _ {\tau}  } \right)
\right|
\lsim 3.4 \cdot 10 ^ { -6 }
\label{Yconstraint}
.
\end{align}
for the yield parameters $ Y _ i \equiv n _ {  i } / s $
with the entropy density  $ s $.
The $ Y _ { X _ \alpha } $ are typically on the order of
$ 10 ^ { -9} \dots 10^{-8} $%
~\cite{Hernandez:2016kel,Ghiglieri:2017csp}.
Since the dominant 
source terms in the kinetic equation for $ L _ { e \rm R } $ are 
the $ X _ \alpha $, we expect $ Y _ { L _ { e \rm R } } $ to be of 
similar size, and~\eqref{Yconstraint} is easily satisfied.

\global\long\def\theequation{\thesection.\arabic{equation}}

\bibliographystyle{jhep}
\bibliography{references}

\end{document}